\documentclass[sigconf,screen,noacm]{acmart}
\AtBeginDocument{%
	}

\setcopyright{rightsretained}
\setcctype{by-nc-nd}
\copyrightyear{2026}

\makeatletter
\def\@copyrightspace{\relax}
\makeatother

\usepackage{tabularray}
\UseTblrLibrary{booktabs}
\usepackage{multirow}
\usepackage{graphicx}
\usepackage{subcaption}
\usepackage{paralist}
\usepackage{enumitem}
\usepackage{xcolor}
\usepackage{xspace}
\usepackage[capitalize, nameinlink]{cleveref}

\usepackage{soul}
\usepackage{marvosym}
\usepackage{placeins}
\usepackage{fontawesome5}

\title{\textsc{Lexplorer}: 
	Navigating~the~Complexity~of~Legal~Document~Landscapes%
}

\newcommand{\paperkeywords}{Legal Technology, Legal Data Science, Visual Analytics, Visualization, Evaluation}

\author{Daniel Fürst}
\authornote{Corresponding author.}
\orcid{0000-0002-0407-2867}
\affiliation{\institution{Data Analysis and Visualization Lab \\ University of Konstanz}
	\city{Konstanz}
	\country{Germany}}
\email{daniel.fuerst@uni-konstanz.de}

\author{Titus Pünder}
\orcid{0009-0009-7779-1924}
\affiliation{\institution{Telos Lab \\ Aalto University}
	\city{Greater Helsinki}
	\country{Finland}}
\email{titus.puender@aalto.fi}

\author{Maximilian T. Fischer}
\authornote{Both authors jointly supervised this work.}
\orcid{0000-0001-8076-1376}
\affiliation{\institution{Data Analysis and Visualization Lab \\ University of Konstanz}
	\city{Konstanz}
	\country{Germany}}
\email{max.fischer@uni-konstanz.de}

\author{Corinna Coupette}
\authornotemark[2]
\orcid{0000-0001-9151-2092}
\affiliation{\institution{Telos Lab \\ Aalto University}
	\city{Greater Helsinki}
	\country{Finland}}
\email{corinna.coupette@aalto.fi}

\date{} %

\definecolor{cb_light_blue}{rgb}{0.651,0.807,0.89}
\definecolor{cb_dark_blue}{rgb}{0.121, 0.47, 0.705}
\definecolor{cb_light_green}{rgb}{0.698, 0.874 ,0.541}
\definecolor{cb_dark_green}{rgb}{0.2, 0.627, 0.172}
\definecolor{cb_light_red}{rgb}{0.984, 0.603, 0.6}
\definecolor{cb_dark_red}{rgb}{0.89, 0.102, 0.109}
\definecolor{cb_light_orange}{rgb}{0.992, 0.749, 0.435}
\definecolor{cb_dark_orange}{rgb}{1, 0.498, 0}
\definecolor{cb_light_purple}{rgb}{0.792, 0.698, 0.839}
\definecolor{cb_dark_purple}{rgb}{0.415, 0.239, 0.603}
\definecolor{cb_yellow}{rgb}{1, 1, 0.6}
\definecolor{cb_brown}{rgb}{0.694, 0.349, 0.157}

\graphicspath{{figures/}}

\crefname{appendix}{appendix}{appendices}
\Crefname{appendix}{Appendix}{Appendices}

\newif\ifshowcomments
\showcommentstrue

\makeatletter
\renewcommand*{\@fnsymbol}[1]{%
	\ifcase #1%
	\or \ensuremath{\ddagger}%
	\or \textasteriskcentered%
	\or \ensuremath{\dagger}%
	\or \ensuremath{\dagger\dagger}%
	\or \ensuremath{\ddagger\ddagger}%
	\or \ensuremath{\mathsection\mathsection}%
	\or \ensuremath{\mathparagraph\mathparagraph}%
	\else
	\@ctrerr
	\fi
}
\makeatother

\makeatletter
\let\papertitle\@title
\makeatother

\newcommand{\lexplorer}{\textsc{Lexplorer}\xspace}

\newcommand{\desaturate}[1]{\textcolor{black!60}{#1}}

\newcommand{\lexTextMode}{{\texttt{T}}\xspace}
\newcommand{\lexDataMode}{{\texttt{D}}\xspace}

\newcommand{\lexOneDoc}{{\desaturate{\small\faEye}}\xspace}
\newcommand{\lexFewDoc}{{\desaturate{\small\faColumns}}\xspace}
\newcommand{\lexManyDoc}{{\desaturate{\small\faFolderOpen}}\xspace}

\newcommand{\lexViewCompare}{{\desaturate{\small\faCodeBranch}}\xspace}
\newcommand{\lexViewRelate}{{\desaturate{\small\faExchange*}}\xspace}

\newcommand{\lexContextPane}{{\desaturate{\small\faWindowMaximize}}\xspace}

\newcommand{\lexFeatureIncoming}{{\desaturate{\small\faFileImport}}\xspace}
\newcommand{\lexFeatureOutgoing}{{\desaturate{\small\faFileExport}}\xspace}
\newcommand{\lexFeatureRibbon}{{\desaturate{\small\faRibbon}}\xspace}
\newcommand{\lexFeatureInlineReference}{{\desaturate{\small\faExternalLink*}}\xspace}
\newcommand{\lexFeatureGlyph}{{\desaturate{\small\faProjectDiagram}}\xspace}
\newcommand{\lexFeatureDossier}{{\desaturate{\small\faBookmark}}\xspace}
\newcommand{\lexFeatureFilter}{{\desaturate{\small\faFilter}}\xspace}
\newcommand{\lexFeatureSearch}{{\desaturate{\small\faSearch}}\xspace}
\newcommand{\lexFeatureReading}{{\desaturate{\small\faReadme}}\xspace}
\newcommand{\lexFeatureToC}{{\desaturate{\small\faList}}\xspace}
\newcommand{\lexFeatureMainText}{{\desaturate{\small\faBars}}\xspace}

\acmDOI{}
\acmISBN{}
\acmConference[]{}%

\begin{document}
	
	\begin{teaserfigure}
	\includegraphics[width=\textwidth]{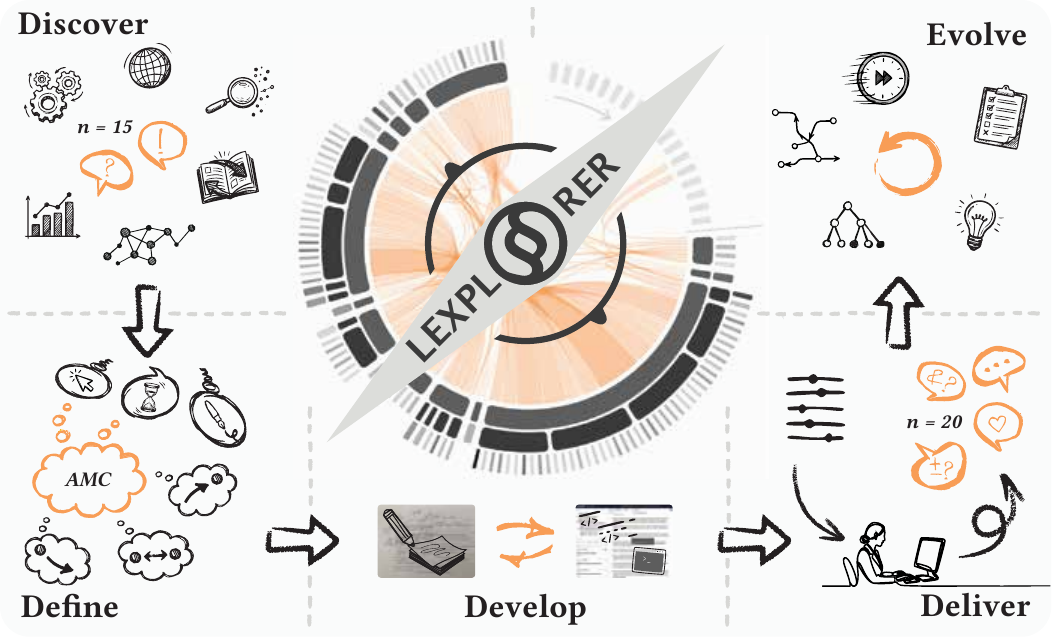}
	\caption{\textbf{\lexplorer embraces legal complexity to support legal work.} Developed through an iterative and ongoing co-design process, \lexplorer offers flexible support for Adaptive Meaning Construction (AMC) in law, using text and visual representations.}
	\Description{The figure shows a workflow process, detailing the five steps in the creations of Lexplorer.
		It starts with a Discovery Phase (interviews with 15 domain experts), over to Define, Develop, Deliver (prototype evaluation with 20 domain experts), and Evolve.} %
	\label{fig:teaser}
	\vspace*{1.5em}
\end{teaserfigure}

	\clearpage
	\begin{abstract}
		As technological and social innovations create novel regulatory challenges, legal systems grow in complexity---%
increasing the need for interfaces that enable effective interactions with legal document collections. 
Through interviews with legal scholars ($n=15$), we find that supporting legal work requires going beyond retrieval-centered legal-information-system paradigms. 
Hence, we propose \lexplorer, a flexible interface for exploring, navigating, and analyzing legal documents, 
based on a taxonomy capturing user intents. 
Distinguishing text and data views for one, few, and many documents, 
\lexplorer enables context-sensitive interactions with evolving collections of interconnected legal texts, 
facilitating \emph{Adaptive Meaning Construction} in law. 
We evaluate \lexplorer with legal scholars ($n=20$) in the context of European Union law, validating our elicited requirements, intent taxonomy, and prototype design. 
Resulting from a close collaboration between visual-analytics researchers and legal scholars, our work also provides nuanced insights into the process required to design interactive systems for expert domains driven by implicit methodological knowledge.

	\end{abstract}



	\maketitle
	
	\fancyhead[LE]{\thepage}%
	\fancyhead[RO]{\thepage}%

	\section{Introduction}\label{sec:introduction}

From the growing capabilities of generative Artificial Intelligence (AI) to the imminent climate crisis, governments around the world often react to global challenges by producing more  and increasingly complex regulation~\cite{katz2020complex,lenz2025democratic}, thereby often creating legal uncertainty that gets gradually reduced via litigation or administrative action~\cite{dari2007uncertainty,fowler2021implement}. 
Legal scholars play a critical role in helping societies integrate the deluge of legal information~\cite{theil2025carefully}, but the interfaces at their disposal offer little support for navigating legal complexity~\cite{furst2025challenges}. 
At the same time, little is known about what interface design could be effective towards supporting their needs and how such a design could be developed in practice. 
\emph{Our work sets out to change this.} 

We investigate how interactive interfaces can support legal work through interviews, prototype design, evaluations, and co-authorship-level collaboration---%
aiming for actionable insights to foster research and support legal scholarship. 
Designing interfaces for the legal domain comes with peculiar challenges:
legal documents and document collections are heterogeneous and constantly evolving, both via amendments of existing documents (e.g., \emph{legislation}) and via the addition of new documents (e.g., \emph{judicial decisions}).
They feature intricate hierarchical, sequential, and reference structures~\cite{coupette2021measuring}---motivating their metaphorical description as \emph{landscapes}. 
Crucially, legal texts are characterized by \emph{intertextuality}~\cite{graham2019intertextuality}, whereby the \emph{meaning} of a text piece depends on its interplay with other pieces of text that may be scattered across different documents. 
This has two implications:
first, \emph{precise} access to original sources remains key, limiting the utility of generative AI.
Second, legal scholars seldom arrive with a question concrete enough to be translated into a fixed set of query results.
Rather, they approach an interface in search of meaning, such that the \emph{relevance} of a text piece often changes dynamically as they encounter additional information. 
These observations suggest that the interaction needs of legal scholars are closer to  \emph{berry-picking}~\cite{batesDesignBrowsingBerrypicking1989} 
than to the traditional view of information retrieval as an iterative query process~\cite{baeza1999modern}.

While legal information systems intend to support \emph{legal work}---%
i.e., work done by people who professionally engage with legal documents, henceforth referred to as \emph{lawyers}---by definition, they are mostly built on the classic paradigm of information retrieval~\cite{van2017concept}. 
For example, \emph{EUR-Lex} and \emph{CURIA} in the European Union (EU), \emph{Westlaw} and \emph{LexisNexis} in the United States of America (USA), or \emph{juris} and \emph{beck-online} in Germany 
readily support keyword search and single-document access, but they currently offer little assistance for tasks like multi-document reading, document-version comparison, or tracing relationships across documents with different semantic roles (e.g., legislation and jurisprudence). 
Prior workflow analyses and interviews with legal practitioners indicate that these limitations force users to rely on tacit knowledge, ad-hoc strategies, manual labor, and a wide variety of tools~\cite{furst2025challenges}.  
The resulting workflows are ripe with media discontinuities and become increasingly fragile as document landscapes grow in complexity~\cite{coupette2023law}. 

Given the scarcity of interfaces that effectively support legal scholars in navigating the complexity of their domain, as well as the lack of insights into how to improve the \emph{status quo}, our work pursues three \textbf{R}esearch \textbf{Q}uestions:
\begin{enumerate}[nosep,label=\bfseries RQ\arabic*]
	\item\label{rq1} What characterizes the \emph{information and interaction needs} of legal scholars? 
	($\to$ \cref{sec:related_work,sec:requirements}) 
	\item\label{rq2} What \emph{interface design} could effectively support these needs? 
	($\to$ \cref{sec:interface-design,sec:prototype-evaluation})
	\item What \emph{development process} is needed to produce high-quality answers to \ref{rq1} and \ref{rq2}? 
	($\to$ \cref{sec:evaluation-design,sec:discussion})
\end{enumerate}

\noindent In answering these questions, we make three \textbf{contributions}:
\begin{enumerate}[nosep,label=\bfseries C\arabic*]
	\item A \textbf{deep understanding of legal~scholars' user needs}  
	based on requirements-elicitation interviews (${n=15}$) and internalized interdisciplinarity (see \ref{c5}), %
	yielding a \textbf{class of problems}, called \emph{Adaptive Meaning Construction} (AMC), to capture the challenges faced by legal scholars in domain-agnostic terms; %
	\item \textbf{\lexplorer}, a \textbf{legal-information-system interface} designed to support Adaptive Meaning Construction in law, %
	tested via a \textbf{detailed prototype evaluation} ($n=20$) in an ecologically validated setup focused on EU law; %
	\item\label{c5} \textbf{Granular insights into the process requirements} for interface design in expert domains, arising from the feedback gathered across our two user studies as well as the co-authorship-level collaboration between legal scholars and visual-analytics experts that led to this work. %
\end{enumerate}
While our work is rooted in the continental legal tradition,  
our interface design readily supports legal document landscapes cultivated in other traditions. 
The principles underlying our interface generalize to any domain where Adaptive Meaning Construction matters, and our design-process learnings will be of interest to anyone building interfaces for domain experts. 

	\section{From Models of Human-Computer Interaction to a Mid-Level, Domain-Agnostic Abstraction}\label{sec:related_work}

As the starting point of our design process, we look to models from Human-Computer Interaction~(HCI) that abstractly describe how people behave when interacting with computer systems for guidance.
For example,~\citeauthor{normanPsychologyEverydayThings1988} observed that interaction follows seven stages, starting with goal formation, followed by three stages of execution, and three stages of evaluation~\cite{normanPsychologyEverydayThings1988}.
Typically, the goal itself unfolds into a hierarchy of sub-goals that reflects categories of intent, where higher-order levels are more abstract and lower-order levels are more specific. 
Since the legal tasks we would like to support typically involve legal information systems, models of information retrieval are particularly relevant. 
These models traditionally assumed the user's goal to be static, modeling user interaction as the repeated querying of an information system until the given information need is satisfied~\cite{baeza1999modern}. 
However, it has been noted that human action is often not planned but rather \emph{opportunistic}, taking advantage of circumstances as they arise~\cite{normanPsychologyEverydayThings1988}---and information seeking is no exception. 

Arguing that the classic model of information retrieval misrepresents human behavior, \citeauthor{batesDesignBrowsingBerrypicking1989} suggests that human search for information is analogous to \emph{berry-picking} in a forest, where berries do not come in bunches but are scattered across bushes~\cite{batesDesignBrowsingBerrypicking1989}.
Later theories of sense-making and analytical reasoning reflect the dynamic nature of berry-picking~\cite{cook2005illuminating, pirolli2005sensemaking, kleinMakingSenseSensemaking2006a}.
On the abstract end of the theory spectrum, in their \emph{Data/Frame theory} of sense-making, \citeauthor{kleinMakingSenseSensemaking2006a} posit that users start with a frame, a viewpoint on their problem, akin to a lens that shapes how they perceive encountered data points, while these data points reciprocally prompt the user to adjust their frame~\cite{kleinMakingSenseSensemaking2006a}.
On the concrete end of the same spectrum, \citeauthor{pirolli2005sensemaking} include an information-foraging loop and a sense-making loop in their domain-specific model tailored to intelligence analysis~\cite{pirolli2005sensemaking}.

While this Data/Frame theory is too abstract to provide guidance for the design of domain-specific interfaces, the model by \citeauthor{pirolli2005sensemaking} is too specific to generalize beyond its highly specialized domain. 
This observation is not specific to these examples but rather \emph{symptomatic}: 
A model gains guidance value as soon as it commits to assumptions about the data, tasks, and environment it addresses, and those commitments are the very ones constraining its generalizability.
The Data/Frame theory transfers broadly because it commits to very little, whereas the information-foraging and sense-making loops guide design because they presuppose the specific corpus, adversarial setting, and reporting duties of intelligence analysis.

The \emph{conflict between specificity and transferability} highlights the \emph{need for a mid-level, domain-agnostic abstraction} that is concrete enough to guide interface design, yet abstract enough to transfer across domains.
Such a layer between the characterization of a domain and the design of encodings and interactions is well-established in visualization~\cite{munznerNestedModelVisualization2009, sedlmair2012design}.
\citeauthor{brehmerMultiLevelTypologyAbstract2013}'s multi-level typology provides a vocabulary for describing tasks at this level in terms of \emph{why}, \emph{how}, and \emph{what}~\cite{brehmerMultiLevelTypologyAbstract2013}, yet the typology classifies tasks that are already known, rather than establishing which tasks arise in a given problem class.
We propose to bridge the remaining gap by committing \emph{not} to a domain but to a \emph{set of properties}, such that the abstraction transfers to any domain sharing the relevant property values. 

To arrive at a mid-level abstraction that can accommodate the specific requirements of legal work, we solicit the expertise of lawyers beyond our interdisciplinary author team, seeking to understand and abstract from how they interact with computer systems to perform legal work~\cite{sedlmair2012design}. 
This leads us to a class of problems we call \emph{Adaptive Meaning Construction}, from which we derive an \emph{intent taxonomy} that serves as the mid-level, domain-agnostic abstraction underpinning our interface design. 
In the next section, we report on our path to that abstraction.

By focusing on \emph{why} and \emph{how} legal information should be made accessible, rather than \emph{what} information, once accessed, will best answer an explicit query, our work complements existing efforts in legal information retrieval~\cite{lauLegalInformationRetrieval2005, saravananImprovingLegalInformation2009, van2017concept, sansoneLegalInformationRetrieval2022}.
As such, it is part of a growing body of literature that investigates the potential of interactive visualization in the legal domain~\cite{lettieriLegalMacroscopeExperimenting2017, lacavaLawNetVizWebbasedSystem2022, resckLegalVisExploringInferring2023, tzanisGraphieNetworkbasedVisual2023}. 
Here, our work pushes the frontier of the field both \emph{methodologically} and \emph{substantively}: 
Following a Double-Diamond-inspired process~\cite{HistoryDoubleDiamond} and building a bridge to research that seeks to understand the impact of technology on legal work \cite{solovey2025interacting,choi2024ai,martinho2025surveying}, 
we develop a full-fledged legal-information-system interface grounded in intents that characterize legal research.

An extended discussion of related work about models of sense-making and visual analytics, Adaptive Meaning Construction in text-heavy domains, and visual analytics for law can be found in \Cref{appendix:extended-related-work}.

\newcommand{\requirementsExpert}[1]{$P_{#1}$\xspace}
\newcommand{\requirementsExpertOne}[0]{\requirementsExpert{1}}
\newcommand{\requirementsExpertTwo}[0]{\requirementsExpert{2}}
\newcommand{\requirementsExpertThree}[0]{\requirementsExpert{3}}
\newcommand{\requirementsExpertFour}[0]{\requirementsExpert{4}}
\newcommand{\requirementsExpertFive}[0]{\requirementsExpert{5}}
\newcommand{\requirementsExpertSix}[0]{\requirementsExpert{6}}
\newcommand{\requirementsExpertSeven}[0]{\requirementsExpert{7}}
\newcommand{\requirementsExpertEight}[0]{\requirementsExpert{8}}
\newcommand{\requirementsExpertNine}[0]{\requirementsExpert{9}}
\newcommand{\requirementsExpertTen}[0]{\requirementsExpert{10}}

\section{Requirements Elicitation and Abstraction}\label{sec:requirements}

Given the limited understanding of the interaction needs experienced by lawyers, 
we conducted semi-structured interviews ($n=15$) to elicit requirements for tool support in understanding and working with legal documents. 
To balance internal and external validity of our insights, 
we required participants to have experience with European Union law. 
This allowed us to include participants educated in countries with different legal cultures, 
while also enabling the translation of our insights into a prototype that could be broadly useful to lawyers operating in the EU. 

In the following, we summarize our participant demographics (\cref{sec:requirements:setup}), 
report our findings (\cref{sec:requirements:findings}), 
and propose an abstraction of these findings to capture the essence of interpretive legal work (\cref{sec:requirements:abstraction}). 
Extended study materials can be found in \Cref{appendix:requirements}.

\subsection{Demographics and Methodology}\label{sec:requirements:setup}

Since the main goal of our work is to support legal scholarship, 
we primarily targeted individuals working in academia as \emph{study participants}.
To add perspective and gauge the potential for generalization to resource-constrained public-interest communities, 
we additionally included individuals working in public institutions and civil-society actors in the scope of our study. 
Our final sample included 10 scholarly, 3 institutional, and 2 societal actors, for a total of 15 participants. 
Scholarly actors were primarily affiliated with 7 different institutions in 3 countries, with core legal education from at least 4 countries, 
including from Northern, Western, and Central Europe. 
One third of our experts identified as female~(5/15), the remaining two thirds identified as male~(10/15). 
At the time of their interview, roughly half of our participants was between 25 and 34 years old~(8/15), the other half between 35 and 54~(7/15). 
Fine-grained participant demographics are reported in \Cref{tab:participant-demographics}.

The \emph{methodology} of our interviews is inspired by the idea of \emph{contextual inquiry}~\cite{beyerContextualDesignDefining1998} as a mixture of conversation and observation, since neither are sufficient on their own~\cite{sedlmair2012design}.
At a high level, we engaged in a conversation with each participant about their legal workflows.
To guide the interview, we followed a list of questions with adaptive probes to capture differences in participants' backgrounds and dig deeper into specific experiences. 
In addition to recording the participant's demographic background, 
each interview consisted of five blocks.
After (1)~capturing a participant's professional context, 
we had them (2)~guide us through a concrete, recent problem in their work, 
before (3)~discussing their general workflows. 
Finally, we (4)~elicited pain points afflicting their work and those parts that already worked for them, before (5)~inquiring into the support they envision. 
All interviews lasted between 45 and 60 minutes. 
They were conducted either in person or via video call, 
recorded, and locally transcribed to facilitate further analysis. 
Participants were recruited via a combination of personal contact networks, mailing lists, and direct cold~outreach and expressed valid consent under applicable laws.

\subsection{Elicited Requirements: Elucidating the How of Legal Work}\label{sec:requirements:findings}

As a basis for our requirements analysis, 
our interviews revealed common workflows and pain points in legal work. 
Most participants described a workflow of \emph{legal research tasks}.
The scope of these tasks is to be understood \emph{broadly}, 
i.e., they are not limited to strictly \emph{academic} research
but rather include every situation in which lawyers need to gather legal information.
At a high level,
lawyers iteratively collect relevant documents
from which they construct \emph{meaning}---%
and
on the basis of which they produce summaries of their findings that constitute their output,
such as a research paper.
To search, participants often use general-purpose search engines like \emph{Google}, rather than querying legal information systems.
The documents they work with are usually provided via a range of specialized databases
that can be loosely distinguished by \emph{jurisdiction} (e.g., EUR-Lex for EU law) and \emph{document type} (e.g., CURIA for materials from EU courts specifically).
Each document is first skimmed
to assess its \emph{relevance}
with regard to the research objective.
The stack of relevant documents grows through consecutive querying 
or as further related documents are identified upon close-reading.
Participants \emph{record} their thought process and findings in text-editing software outside of the legal information system.
In understanding legal documents, they are supported by the document structure
and benefit from tacit knowledge and prior experience.
Their work ends when all relevant documents have been processed---%
or, as completeness can often not be guaranteed, when their result is \emph{socially complete}.

Our study participants reported a range of \emph{common pain points}.
With regard to \emph{data},
they highlighted the subpar quality of search results
and the problem of missing documents,
which has led to a general lack of trust
in the completeness and correctness of legal information systems.
When it comes to their actual \emph{workflows},
participants wished for easier navigation between documents
to reduce context-switching costs. 
Instead of manually rotating between multiple PDFs, browser tabs, and text search, 
they would appreciate one comprehensive system
that provides interlinked documents also across document types.

From the common workflows and pain points sketched above, we distill \textbf{R}equirement \textbf{G}roups. 
A more detailed description of these groups, 
including which participants raised which points, 
can be found in \Cref{appendix:extended-requirements}. 
In a nutshell, 
to navigate legal document landscapes, 
lawyers need: 
\vspace*{0.5cm} %
\begin{enumerate}[nosep,label=\bfseries RG\arabic*]
	\item\label{rg:data-access} \textbf{Data Access}---quick and easy retrieval, providing an expressive data response
	that includes metadata, temporal context, as well as intra-document and inter-document relationships.
	\item\label{rg:search-and-navigation} \textbf{Search and Navigation}---semantic search, fine-grained control over search parameters,
and linked-document navigation.
	\item\label{rg:versioning-and-comparison} \textbf{Versioning and Comparison}---the ability to view the history of a document, detect changes between different document versions, and track the evolution of concepts over time.
	\item\label{rg:relationships-and-context} \textbf{Relationships and Context}---relationships between documents made visible as metadata
and accessible via hyperlinks.
	\item\label{rg:summarization-and-visualization} \textbf{Summarization and Visualization}---document comprehension supported by simplified descriptions, statistical information, or visual representations (such as hierarchies, timelines, and networks). 
	\item\label{rg:usability-and-performance} \textbf{Usability and Performance}---an intuitive, learnable interface capable of mitigating information overload and reducing real work. 
\end{enumerate}
Participants expressed broad dissatisfaction with existing interfaces across all groups of requirements.

\begin{figure*}[t]
	\centering
	\includegraphics[width=\textwidth]{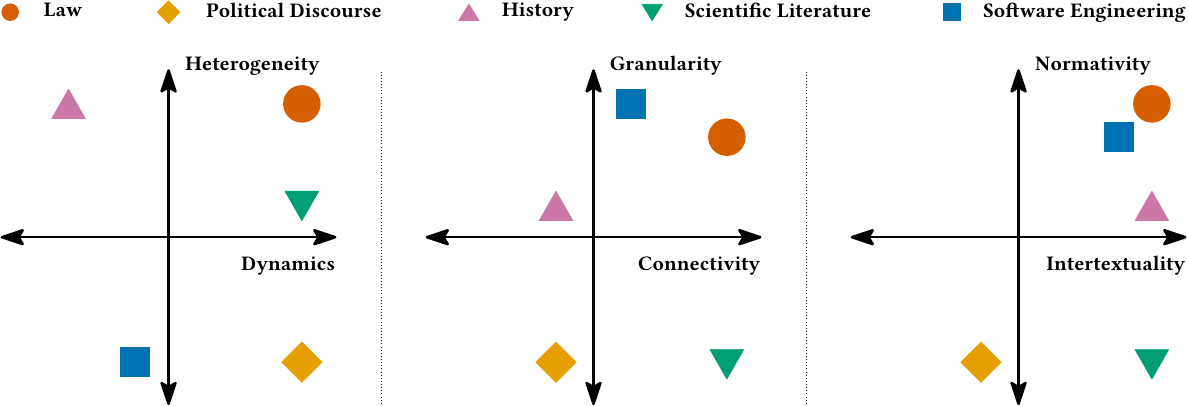}
	\begin{subfigure}{0.333\linewidth}
		\subcaption{Heterogeneous Dynamics}
	\end{subfigure}~%
	\begin{subfigure}{0.333\linewidth}
		\subcaption{Granular Connectivity}
	\end{subfigure}~%
	\begin{subfigure}{0.333\linewidth}
		\subcaption{Normative Intertextuality}
	\end{subfigure}
	\caption{\textbf{Legal texts exhibit a unique combination of properties.} 
		Legal texts feature a special combination of heterogeneous dynamics, granular connectivity, and normative intertextuality (see \Cref{appendix:legal-work}). 
		This separates the requirements of legal work from those found in other text-heavy domains, 
		which are similar to law in  \emph{some} but not \emph{all} dimensions. 
		The position of all domains on the coordinate systems should be read heuristically, i.e., as indicating a tendency of a specific domain toward a particular area in the coordinate system, rather than a precise position. 
	}\label{fig:legal-work}
	\Description{The figure shows three scales for the three properties of domains captured by Adaptive Meaning Construction: heterogeneous dynamics, granular connectivity, and normative intertextuality. The figure places law, political discourse, history, scientific literature, and software engineering on these axes.} %
\end{figure*}

\subsection{Abstraction of Requirements: Moving from the How to the Why of Legal Work}\label{sec:requirements:abstraction}

The requirements elicited from our study participants focus on \emph{how} lawyers would like to interact with legal information systems (\emph{method})
without capturing \emph{why} they would like to do so. 
While echoing previous findings from the literature \cite{furst2025challenges}, 
method-focused requirements offer little guidance for interface design. 
Rather, one could argue that \ref{rg:data-access} and \ref{rg:usability-and-performance} are prerequisites for \emph{any} information system, 
and that \ref{rg:search-and-navigation}--\ref{rg:summarization-and-visualization} are equally applicable to other domains characterized by dynamic corpora of interrelated texts, 
such as political discourse analysis \cite{sperrleVIANAVisualInteractive2019}, 
history~\cite{scheirerSenseConnectionAutomatic2016}, 
scientific literature exploration \cite{duckFindingNeedlesDocument2025}, 
or 
software engineering \cite{horvathUsingAnnotationsSensemaking2022}. 
To understand the extent to which interface-design insights from such domains could be translated to the legal domain, 
and to derive actionable guidance for interface design, 
we need to look beyond the \emph{How} of legal work to its \emph{Why} (\emph{motivation}). 

At its core, all text-heavy work involves establishing \emph{meaning}. 
In the legal domain, 
the target of this endeavor can be, e.g., a specific legal text, rule, or concept. 
Legal meaning determination is particularly challenging 
because legal texts offer a peculiar combination of characteristics that arise from the organization of legal systems (see \Cref{appendix:legal-work} for a primer).
In particular, legal corpora exhibit the following \textbf{P}roperties:
\begin{enumerate}[nosep,label=\bfseries P\arabic*]
	\item\label{p1} \textbf{Heterogeneous dynamics}---%
	Legal texts and legal meaning can change (\emph{dynamics}), 
	and they can do so in different ways (\emph{heterogeneity}) depending on the actors and document types involved. 
	For example, legislation is often edited `in place' via amendments, whereas case law mostly changes via growth (i.e., the addition of new decisions to the corpus). 
	\item\label{p2} \textbf{Granular connectivity}---%
	Legal texts are linked via intricate explicit and implicit relationships (\emph{connectivity}), 
	and these relationships occur at a very fine level of resolution (\emph{granularity}). 
	For example, a paragraph of a judicial decision may reference paragraphs of several other judicial decisions, along with pieces of legislation cited two levels below their `street-level address', 
	where the granular reference carries the crucial meaning (e.g., Article 6(1)(a) of the General Data Protection Regulation [GDPR] and Article 6(1)(f) GDPR have very different implications).
	\item\label{p3}\textbf{Normative intertextuality}---%
	The meaning of a piece of legal text depends on its interplay with other pieces of legal text (\emph{intertextuality}), 
	and that influence can determine what interpretations of a text are valid (\emph{normativity}). 
	For example, 
	a ruling by the European Court of Justice can authoritatively shape what real-life scenarios are considered as concerning `personal data' under the GDPR. 
\end{enumerate}

Since each property combines two features that can be realized to different degrees, 
the properties sketched above can also be thought of as marking the extreme positions of legal text on three coordinate systems. 
As illustrated in \Cref{fig:legal-work}, 
together, these positions capture essential differences between law and other text-heavy domains. 
However, these domains are united in the relevance of meaning determination---%
i.e., they share some of their \emph{Why}. 

We refer to the broad class of problems shared by text-heavy domains as \emph{Adaptive Meaning Construction}~(AMC). 
Rather than being tied to any specific domain, AMC represents an umbrella concept to characterize problems in domains that feature different configurations of a shared set of properties. 
Domain-agnostic, yet configuration-specific interface-design guidance, then, can be derived from locating the individual domains on the coordinate systems defined by \ref{p1} to \ref{p3} 
and deriving corresponding \emph{analytic intents}. 
For example, 
practitioners in a highly \emph{dynamic} domain will frequently need to understand how an external change (\emph{event trigger}) affects meaning---%
from the perspective of the text carrying the change (\emph{outgoing impact determination}), 
a different text, rule, or concept they care about (\emph{incoming impact determination}), 
or to adjust their mental model based on a previous version in light of a new version (\emph{comparison}). 
Similarly, 
a practitioner in a highly \emph{intertextual} domain will regularly be motivated to determine the meaning boundaries of an idea as part of their work (\emph{task trigger}), 
be it because the idea itself is of interest (\emph{definition}),  
because they need to decide whether an example is an instance of the idea (\emph{classification}), 
or because the meaning of one idea can only be understood in its interplay with the meaning of some other idea (\emph{relation}). 

\begin{table*}[t]
	\centering
	\caption{\textbf{A two-dimensional taxonomy of intents can guide interface design for AMC problems.} 
		We show how intents derived from Adaptive Meaning Construction (left) are mapped to design elements of the \lexplorer interface (right), adding example questions for each specific intent.
	}\label{tab:intent-taxonomy}
	
	\begin{tabular}{lrl|ll}
		\toprule
		\multicolumn{3}{l}{\bfseries Adaptive-Meaning-Construction Intents}&\multicolumn{2}{|l}{\bfseries \lexplorer Interface Support}\\
		Trigger&Voice&Intent&Relevant Views&Relevant Features\\
		\midrule
		Event-Based& Active&\emph{Outgoing Influence Determination}&\lexOneDoc$\mid$ \lexManyDoc $\times$ \lexTextMode$\mid$ \lexDataMode &\lexFeatureMainText~\lexFeatureReading~\lexFeatureInlineReference~\lexFeatureOutgoing~\lexFeatureGlyph\\
		&&\multicolumn{3}{l}{Example: How does this new text affect the meaning of other~texts?}\\\cmidrule{2-5}
		& Middle&\emph{Comparison}&\lexFewDoc~\lexTextMode&\lexFeatureMainText~\lexFeatureToC~\lexViewCompare\\
		&&\multicolumn{3}{l}{Example: How does the meaning of this new text differ from that of previous texts?}\\\cmidrule{2-5}
		& Passive&\emph{Incoming Influence Determination}&\lexOneDoc$\mid$ \lexManyDoc $\times$ \lexTextMode$\mid$ \lexDataMode&\lexFeatureMainText~\lexFeatureReading~\lexFeatureIncoming~\lexFeatureRibbon\\
		&&\multicolumn{3}{l}{Example: How is the meaning of the present text affected by this new~text?}\\
		\midrule
		Task-Based &Active&\emph{Definition}&\lexOneDoc$\mid$ \lexManyDoc $\times$ \lexTextMode$\mid$ \lexDataMode&\lexFeatureMainText~\lexFeatureFilter~\lexFeatureInlineReference~\lexFeatureRibbon~\lexFeatureDossier\\
		&&\multicolumn{3}{l}{Example: What is the meaning of concept $X$?}\\\cmidrule{2-5}
		& Middle&\emph{Relation}&\lexFewDoc~\lexDataMode&\lexFeatureMainText~\lexFeatureToC~\lexViewRelate~\lexFeatureGlyph \\
		&&\multicolumn{3}{l}{Example: How does this concept interact with this other concept?}\\\cmidrule{2-5}
		& Passive&\emph{Classification}&\lexOneDoc$\mid$ \lexManyDoc $\times$ \lexTextMode$\mid$ \lexDataMode&\lexFeatureMainText~\lexFeatureSearch~\lexFeatureRibbon~\lexFeatureGlyph\\
		&&\multicolumn{3}{l}{Example: Is example $a$ included in the meaning of concept~$X$?}\\
		\bottomrule
	\end{tabular}
	\Description{%
		The table shows how intents derived from Adaptive Meaning Construction are mapped to design elements of the Lexplorer interface, adding example questions for each specific intent. 
		From left to right, the first three columns describe Adaptive-Meaning-Construction Intents, taxonomizing them by trigger (event-based vs. task-based) and voice (active, middle, or passive).
		The last two columns describe how Lexplorer supports the AMC intents in question, using icons to denote the relevant view types, view modes, and features.
	}
\end{table*}

Therefore, as summarized in the left part of \Cref{tab:intent-taxonomy}, 
AMC intents can be organized into a two-dimensional taxonomy, 
grouped first by the nature of their \emph{trigger} (event-based vs. task-based), 
and then by their \emph{voice} of meaning construction (active vs. middle vs. passive, borrowing a metaphor from linguistics). 
However, 
the detailed design of an interface aiming to support these intents will depend also on the other dimension of the coordinate systems underlying the distinctions. 
Additionally, \emph{granular connectivity} sets the intensity and resolution at which all intents operate, 
guiding prioritization as well as suggesting suitable views and features.
In the next section, we leverage this insight to guide the design of a legal information system that accommodates the unique AMC configuration of the legal domain.

	\section{Interface Design}\label{sec:interface-design}

From law's position in the coordinate systems shown in \Cref{fig:legal-work}, 
we can deduce three design implications for interfaces aiming to support Adaptive Meaning Construction in the legal domain. 
First, 
while the familiar \emph{one-document view} (\lexOneDoc) remains crucial for focused reading, 
due to \emph{heterogeneous dynamics}, 
support for \emph{comparing} and \emph{relating} documents should be natively available, e.g., via a \emph{few-document view} (\lexFewDoc). 
Second, due to \emph{normative intertextuality}, 
an interface should expose relationships between documents in both directions, 
enabling users to determine not just the \emph{outgoing influence} (\lexFeatureOutgoing) of a text in focus but also its \emph{incoming influence} (\lexFeatureIncoming). 
Moreover, authoritative legal meaning will often only be understandable by jointly considering multiple text passages (\lexFeatureMainText) from differentially important documents in parallel, 
suggesting contextualized multi-document reading (\lexFeatureReading) as a useful capability to support \emph{definition} and \emph{classification} intents, contextual quick lookups in a side pane (\lexContextPane),
as well as a dedicated \emph{many-document view} (\lexManyDoc) to allow overview and discovery of potential meaning interactions beyond pairs of documents via \emph{node-link diagrams} (\lexFeatureGlyph).
And third, given \emph{granular connectivity}, 
reference exploration, filtering, and navigation should be enabled at the finest possible resolution. 
Here, one can take inspiration from existing designs in software engineering, as this domain is close in the coordinate system defined by the relevant features. 
This suggests, for example, 
exposing the hierarchical structure of a document via an interactive table of contents (ToC, \lexFeatureToC) 
and using in-(con)text \emph{ribbons} (\lexFeatureRibbon) to annotate relationships.

To act on the implications delineated above, 
we propose \lexplorer, an interface supporting AMC in legal work. 
While our initial prototype development is focused on EU Law, 
the process and patterns we employ do not commit to any particular legal system. 
Thus, they should be equally applicable in other jurisdictions (see also \Cref{sec:discussion}). 

\paragraph{Design Process}
Our design process was highly iterative and interdisciplinary,  
with legal scholars co-developing the interface alongside visual-analytics researchers. 
While details on our design process are collected in \Cref{appendix:design-process}, 
in a nutshell, 
we started by collecting inspirational visualization from related work in a mood board.
In parallel, we started sketching first ideas on analog media.
Inspired by what has been called \emph{Pixar Planning}~\cite{flyvbjerg2023big}, our goal was to map out the set of possible interactions between the views of the interface in meticulous detail.
We continuously expanded the collection of design ideas without committing to any design early on, allowing the ongoing interviews to reshape, rather than confirm, our design space.
In parallel, we started setting up the technical infrastructure of the project. 
Once we had concluded our requirements-elicitation interviews and analyzed their findings, 
we started rapid prototyping of the interface. 
While we handed the burden of writing the frontend code to AI, 
we closely monitored the correct implementation of intents with internal feedback loops.

\begin{figure*}[t]
	\centering
	\begin{subfigure}{0.475\linewidth}
		\centering
		\includegraphics[width=\linewidth]{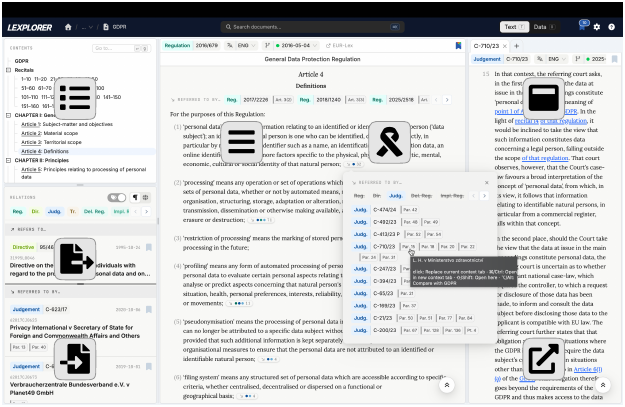}
		\subcaption{Text Mode}\label{fig:one-document-view:text-mode}
	\end{subfigure}~%
	\begin{subfigure}{0.475\linewidth}
		\centering
		\includegraphics[width=\linewidth]{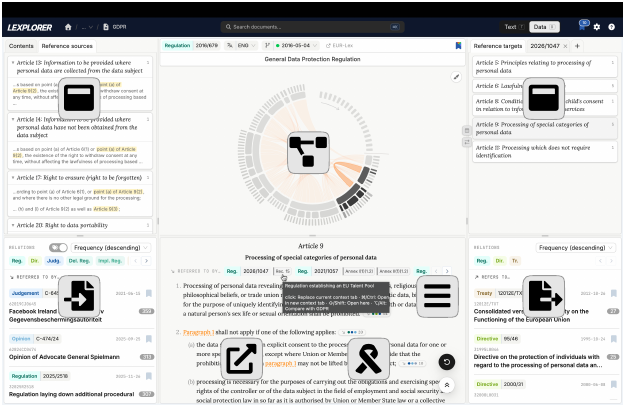}
		\subcaption{Data Mode}\label{fig:one-document-view:data-mode}
	\end{subfigure}
	
	\caption{\textbf{One-document views.}
		Our one-document views support users in reading and contextualizing a single document.
		In Text Mode (\subref{fig:one-document-view:text-mode}), the full text of a legal document is shown in a larger \lexFeatureMainText center pane, in our example case the GDPR, together with several side panes and in-context information.
		Interaction is possible through keyboard shortcuts, multiple context menus, and clicking.
		Navigation is aided by a multi-level \lexFeatureToC ToC in a top left pane, while both \lexFeatureIncoming incoming and \lexFeatureOutgoing outgoing references are displayed in a (by default contextual) bottom left pane.
		The \lexFeatureMainText main text in the center features metadata on top, the main text, inline outgoing references inline \lexFeatureInlineReference and incoming references as ribbons \lexFeatureRibbon, 
		with mini-previews to also explore both directions directly within the paragraph or as popup.
		On the right side, a \lexContextPane contextual pane offers quick previews of selected or clicked documents for contextual understanding.
		In Data Mode (\subref{fig:one-document-view:data-mode}), the \lexFeatureMainText main text is retained but shifted down to be joined by a radial \lexFeatureGlyph glyph representation that visually shows the connections between individual paragraphs in the document.
		When selecting a specific paragraph in the hierarchical radial \lexFeatureGlyph glyph, which reflects the ToC structure, the connections are highlighted, while the top left and top right side panes provide additional snippets of reference sources and targets.
		In our example, the screenshot shows which other paragraphs refer to Article 9 of the GDPR.}
	\label{fig:one-document-view}
	\Description{The figure shows two screenshots of Lexplorer for the one-document views, displaying their text mode and data mode.
	These views support users in reading and contextualizing a single document.
	In Text Mode (a), the full text of a legal document in shown in a larger center pane, in our example case the GDPR, together with several side panes and in-context information.
	Interaction is possible through keyboard shortcuts, multiple context menus, and clicking.
	Navigation is aided by a multi-level ToC in a top left pane, while both incoming and outgoing references are displayed in a (by default contextual) bottom left pane.
	The main text in the center features metadata on top, the main text, and inline references with mini-previews to also explore both directions directly within the paragraph or as popup.
	On the right side, a contextual pane offers quick previews of selected or clicked documents for contextual understanding.
	In Data Mode (b), the main text is retained but shifted down to be replaced by a radial glyph representation that visually shows the connections between individual paragraphs in the document.
	When selecting a specific paragraph in the hierarchical radial glpyh, reflecting the ToC structure, the connections are highlighted, while the top left and top right side panes provide additional snippets of the source and target of references.
	In our example, the screenshot shows which other paragraphs refer to Article 9 of the GDPR.} %
\end{figure*}

\begin{figure*}[t]
	\centering
	\begin{subfigure}{0.475\linewidth}
		\centering
		\includegraphics[width=\linewidth]{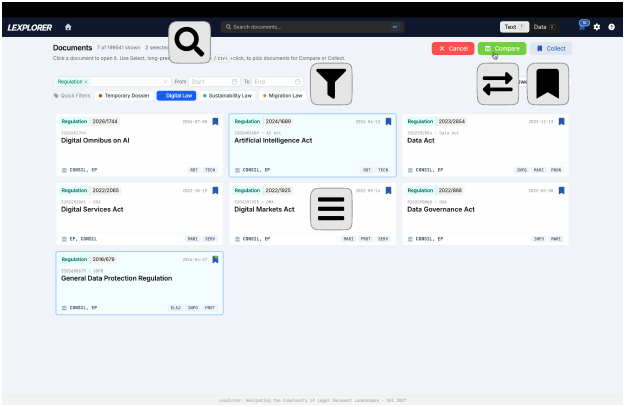}
		\subcaption{Text Mode}\label{fig:many-documents-view:text-mode}
	\end{subfigure}~%
	\begin{subfigure}{0.475\linewidth}
		\centering
		\includegraphics[width=\linewidth]{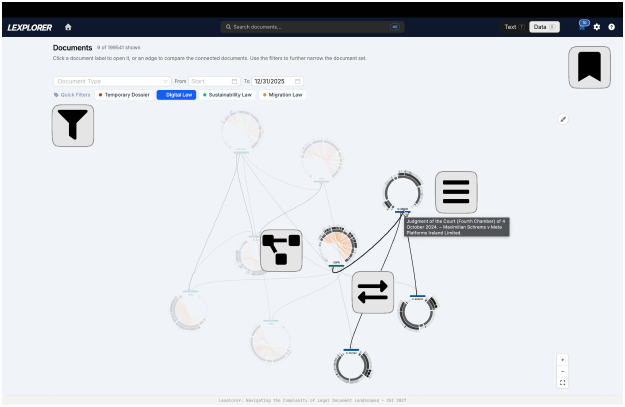}
		\subcaption{Data Mode}\label{fig:many-documents-view:data-mode}
	\end{subfigure}
	
	\caption{\textbf{Many-document views.}
		Our many-document views focus on understanding relations between a small- to medium-sized collection of documents and, for the text mode, double as our landing page.
		In Text Mode (\subref{fig:many-documents-view:text-mode}), the main focus lies on accessing or selecting works that are already known. 
		Documents are presented as cards showing essential information such as short and long title, date, document type, unique identifier, authoring bodies, and keyword tags.
		These can then be individually opened for \lexFeatureMainText reading or batch-selected for \lexViewRelate comparison.
		To narrow the scope of documents, users can combine \lexFeatureFilter metadata filters on document attributes with both instant meta-data-based search as well as a \lexFeatureSearch full-text search, supporting both targeted retrieval and broader exploratory filtering.
		Users can collect documents they want to investigate together in \lexFeatureDossier dossiers (shown is a dossier containing documents related to \emph{Digital Law}).
		In Data Mode (\subref{fig:many-documents-view:data-mode}), the filtered set of documents is shown as a force-directed layout of \lexFeatureGlyph glyphs that are \lexViewRelate connected for understanding their interrelations.}
	\label{fig:many-documents-view}
	\Description{The figure shows two screenshots of Lexplorer for the many-document views, displaying their text mode and data mode.
	These views focus on understanding relations between a small- to medium-sized collection of documents and, for the text mode, double as main start page.
	In Text Mode (a), the main focus lies on accessing or selecting already known works.
	Documents are presented by cards showing essential information such as short and long title, date, document type, unique identifier, authoring bodies, and keyword tags.
	These can the be individually opened for reading or batch-selected for comparison.
	To narrow the scope of documents, users can combine metadata filters on document attributes with both instant meta-data-based search as well as a full-text search, supporting both targeted retrieval and broader exploratory filtering.
	Users can collect documents in dossiers (shown is the set of documents collected into the crafted dossier Digital Law) as sets of sources that address a legal inquiry or specific sub-questions.
	In Data Mode (b), the filtered set of documents is shown as a force-directed layout of glyphs, and connections between documents can be clicked to access the detailed comparison mode.} %
\end{figure*}

\begin{figure*}[t]
	\centering
	\begin{subfigure}{0.475\linewidth}
		\centering
		\includegraphics[width=\linewidth]{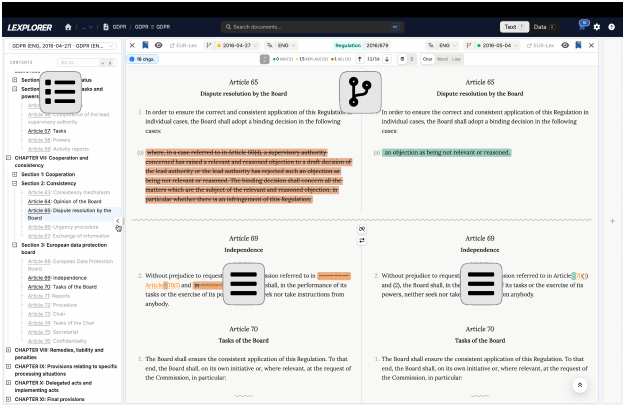}
		\subcaption{Two Documents}\label{fig:compare-view:text-mode:two-documents}
	\end{subfigure}~%
	\begin{subfigure}{0.475\linewidth}
		\centering
		\includegraphics[width=\linewidth]{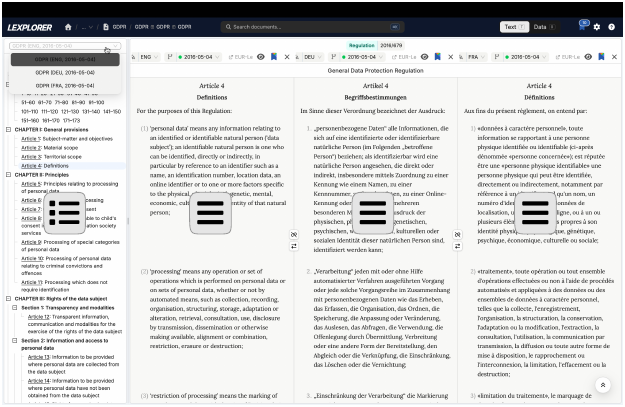}
		\subcaption{Three Documents}\label{fig:compare-view:text-mode:three-documents}
	\end{subfigure}
	
	\caption{\textbf{Few-document view in text mode.}
		The few-document text view focuses on understanding \lexViewCompare changes between document versions or differences between documents.
		For example, one can track changes across different temporal versions of the same document (\subref{fig:compare-view:text-mode:two-documents}) or investigate differences between linguistic versions via parallel reading (\subref{fig:compare-view:text-mode:three-documents}).
		The left pane features again a \lexFeatureToC ToC for navigation, while the main panes show typically two to three individual \lexFeatureMainText main texts.
		For the comparison, statistics and different granular \lexViewCompare difference computation options are available, like side-by-side (shown) or inline, the granularity level, and folding.}
	\label{fig:compare-view:text-mode}
	\Description{The figure shows two screenshots of Lexplorer for the few-document view in text mode.
	The text-focused few-document view focuses on understanding changes between document versions or difference between documents.
	For example, one can compare how a law has been revised (a) or understand differences between translated versions for parallel reading (b).
	The left pane features again a ToC for navigation, while the main panes show typically two to three individual main texts.
	For the comparison, statistics and different granular difference computation options are available, such as side-by-side (shown) or inline, the granularity level, and folding.} %
\end{figure*}

\begin{figure*}[t]
	\centering
	\begin{subfigure}{0.475\linewidth}
		\centering
		\includegraphics[width=\linewidth]{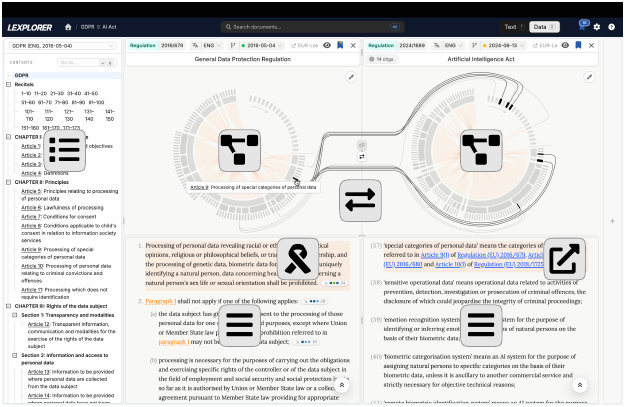}
		\subcaption{Two Documents}\label{fig:compare-view:data-mode:two-documents}
	\end{subfigure}~%
	\begin{subfigure}{0.475\linewidth}
		\centering
		\includegraphics[width=\linewidth]{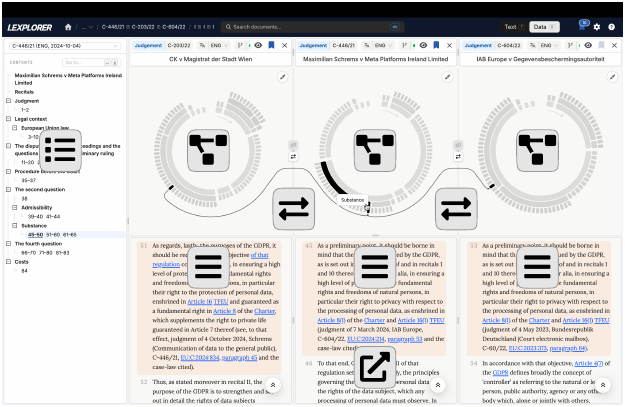}
		\subcaption{Three Documents}\label{fig:compare-view:data-mode:three-documents}
	\end{subfigure}
	
	\caption{\textbf{Few-document view in data mode.}
	The few-document data view extends the text-based comparison of the few-document view (see also \Cref{fig:compare-view:text-mode}) by providing a \lexFeatureGlyph glyph-based representation.
	However, in addition to relations \emph{within} one and the same document, article-level \lexViewRelate relations \emph{between} the selected documents are also shown, facilitating \lexFeatureInlineReference linked-document navigation and mental mapping.
	In our example, the glyphs in the few-document view (\subref{fig:compare-view:data-mode:two-documents}) show which Articles of the EU AI Act (right \lexFeatureGlyph glyph and \lexFeatureMainText text) refer to Article 9 of the GDPR (left \lexFeatureGlyph glyph and \lexFeatureMainText text).
	For three documents (\subref{fig:compare-view:data-mode:three-documents}), 
	the positioning becomes important, as the glyph in the center pane shows the relations to either document on each side.
	In the case shown, the positioning reveals a citation chain within the jurisprudence of the European Court of Justice. 
	}
	\label{fig:compare-view:data-mode}
	\Description{The figure shows two screenshots of Lexplorer for the few-document view in data mode.
	This mode extends the text-based comparison of the few document view by providing a glyph-based representation.
	However, instead of showing intra-relations, inter-relations between the selected documents are shown, i.e., which paragraphs in particular refer to (or are referred by) the corresponding documents to support mental mapping.
	In our example, the glyphs in the two document view (a) show which paragraphs of the EU AI Act (right glyph and text) refer to Article 9 of the GDPR (left glyph and text).
	For more than two documents, the positioning becomes important, as the center glyph shows the relations to either document on each side.} %
\end{figure*}

\paragraph{Overview of \lexplorer}
In the following, we provide a concise overview of our \lexplorer interface, 
illustrated with annotated screenshots (\Cref{fig:one-document-view,fig:compare-view:data-mode,fig:compare-view:text-mode,fig:many-documents-view}). 
Further information on the interface features, including more details on the rationale behind individual view types and view modes, 
can be found in \Cref{appendix:interface-details}.

\lexplorer is designed as a React-based web-application interface. 
As elaborated in the right part of \Cref{tab:intent-taxonomy}, 
it implements the intent taxonomy derived in \Cref{sec:requirements:abstraction} by leveraging a set of coordinated one-, few-, and many-document views over two different modes (\emph{text} and \emph{data}). 
The two modes of each view focus stages of intent realization:
The \emph{text mode} supports a view's corresponding intents \emph{directly} when the relevant textual content has already been identified, 
whereas the \emph{data mode} provides context via structural visualizations and relational statistics attuned to intents, 
allowing users to identify contextually relevant content and thus \emph{indirectly} supporting intent realization. 
All views are linked through shared selection and highlighting, which ensures that moving between intents preserves context, 
avoiding costly context switching.

The overall interface features a top menu bar that contains navigational elements, a context-sensitive search bar, text- and data-mode switches, bookmark folders (called \emph{dossiers}), and settings.
The main area depends on the view, but it typically features one or more central elements and contextual information on multiple collapsible side panes.  
In its default text mode, 
the \emph{one-document view} supports contextualized close-reading of legal documents, 
helping lawyers understand \emph{incoming} and \emph{outgoing} influences between texts as well as discover the information needed to \emph{define} concepts or \emph{classify} individual instances. 
While the one-document text view resembles traditional text-based interfaces,
as shown in \Cref{fig:one-document-view}, 
\lexplorer goes beyond simple close reading: 
Already in text mode, it offers fine-grained access to and in-context skimming of incoming and outgoing references.
In data mode, the one-document view additionally offers prioritized access to related documents and within-document interactions between texts. 

The landing page of \lexplorer is the \emph{many-document view}, 
depicted in \Cref{fig:many-documents-view}. 
In text mode (default), this view provides an overview of legal sources available, 
offering selection, sorting, filters, as well as keyword and full-text search.
In data mode, 
the many-document view exposes the connectivity between documents held in the active folder, 
enabling easy exploration of interactions within user-defined groups of documents. 
Conceptually located between the one-document view and the many-document view,  
the \emph{few-document view}, depicted in \Cref{fig:compare-view:text-mode} and \Cref{fig:compare-view:data-mode}, 
features fine-grained difference computations in the style of \emph{git} for different \emph{temporal versions} of the same document, 
linked reading for different \emph{linguistic versions} of the same document, 
and visually navigable relationship exploration for two or three \emph{different documents}. 

The \lexplorer frontend is paired with a preprocessing and computing backend (Python with FastAPI and Postgres) that handles an intricate data-ingestion pipeline (from parsing documents in heterogeneous formats into a standardized representation via metadata extraction and reference extraction to data storage and indexing) as well as difference computations, statistics generation, and text search. 
At the time of prototype evaluation (see \Cref{sec:prototype-evaluation}), 
\lexplorer enabled full-text access, navigation, and relationship exploration for around 200k EU legal documents. 
Once backend data ingestion is completed, it will offer at least the coverage provided by the official EU legal data portal (EUR-Lex)---%
a crucial feature for our target user group (see \Cref{sec:discussion}). 

	\section{Evaluation Design}\label{sec:evaluation-design}

Having derived our interface design from Adaptive Meaning Construction as our mid-level abstraction (see \Cref{sec:interface-design}), 
the natural next step in our work was to expose the resulting \lexplorer prototype to rigorous evaluation. 
Ideally, this evaluation would not just validate our general approach. 
Rather, it would also allow us to improve our prototype going forward 
and offer higher-level insights into best practices for interface design supporting AMC in the legal domain.   
Therefore, we aimed to design our evaluation around three goals:
\begin{enumerate}[label=\bfseries G\arabic*,nosep]
	\item\label{g1} Ensure that the current prototype is based on \emph{valid assumptions}, i.e., 
	that its \emph{functionalities} support realistic legal tasks 
	and that its \emph{interaction patterns} correspond to thinking patterns involved in AMC. 
	\item\label{g2} Assess the usability and effectiveness of the specific \emph{design choices} made and elicit \emph{general feedback} on the prototype \emph{implementation}.
	\item\label{g3} Gather \emph{feature requests} on the prototype to guide its future development. 
\end{enumerate}

\paragraph{Evaluation-Design Process to Ensure Ecological Validity}
There was a lack of consensus in the author team about what type of evaluation would support the above-stated goals. 
Roughly, the group was split between legal scholars, 
who favored a \emph{participatory evaluation}, ideally giving lawyers ample time to interact with the prototype and use it in their own work, 
and visual-analytics experts, 
who favored a \emph{quantitative evaluation} to ensure objectivity. 
Since the main artifact to be evaluated seemed closer to the expertise of the latter group, 
the legal scholars initially deferred to the visual-analytics experts, 
leading to an initial draft for a quantitative evaluation. 
This draft was the result of a rather difficult process whereby the lawyers on the team translated results from their usual workflows into quantifiable artifacts. 
However, this evaluation design completely failed a test for ecological validity with a legal scholar not involved in the study: 
Our first test subject hardly interacted with the prototype at all, 
constantly asking how the task would be relevant for their work. 
The group then pivoted to a \emph{qualitative evaluation} designed to strike a compromise between the conflicting positions, 
formulating a fixed set of open-ended tasks without expecting users to produce specific pieces of information as results.
The revised draft evaluation then passed a renewed test for ecological validity: 
Our second test subject naturally engaged in interactions with the prototype, 
repeatedly expressing that what they were doing resembled not only their own workflows but also that of their colleagues. 
Therefore, we decided to move forward with the qualitative approach, 
keeping participatory evaluation for the next version of our prototype. 
A detailed breakdown of the questions and tasks included in the study design can be found in \Cref{appendix:evaluation-structure}.

\paragraph{Similarities between Quantitative and Qualitative Designs}

Despite their radically different ecological-validity outcomes, our quantitative and qualitative designs shared most evaluation components.  
Following introductory formalities, 
each user was prompted to answer a range of questions about what they \emph{expect} from and \emph{appreciate} in a legal information system.
This step was primarily designed to validate the requirements elicited in our first expert study, catering to \ref{g1}. 
After demonstrating both EUR-Lex (the official gateway to European Union law serving as the baseline) and our prototype through \emph{guided tours}, 
participants were asked to \emph{interact with the prototype} to work on three legal tasks, directly gathering information on \ref{g2} as well as indirect information on \ref{g1} and \ref{g3}. 
These tasks were embedded in distinct legal fields: 
Digital Law, Sustainability Law, and Migration Law.
Each task was designed to cover a different perspective on the document landscape,
namely leveraging cross-references,
comparing temporal versions of a single document,
and
exploring a set of multiple interconnected legal documents. 
We concluded our user study with a feedback session. 
This session elicited \emph{quantitative feedback} inquiring about the usability of the prototype via Likert Scales inspired by SUS and NASA-TLX, 
informing \ref{g2},
as well as \emph{qualitative feedback} via a semi-structured interview
that further explored a participant's interaction experience and elucidated potential misalignments between existing and desired functionality, supporting \ref{g3}.

\paragraph{Differences between Quantitative and Qualitative Designs}
The only difference between the quantitative and the qualitative design was the way in which the legal tasks were posed.
In the \emph{quantitative setting}, 
the tasks asked users to provide a fixed number of identifiers for specific (parts of) legal documents that met certain objectively verifiable requirements, 
contextually embedded in the legal fields shared by both evaluation designs.
Users completed each set of tasks twice, 
once using the baseline system (EUR-Lex) and once using \lexplorer (with system order randomized across study participants).
In the \emph{qualitative setting}, 
the first task asked users to explore how a specific definition in a legislative act---%
i.e., the definition of `personal data' in the \emph{General Data Protection Regulation}---%
had been developed through jurisprudence.
The second task required users to identify key changes over time in a prominent legal document, the \emph{European Climate Law}, 
and the third task encouraged them to explore a legal field shaped by a set of closely related documents, the \emph{Common European Asylum System}. 
Each task was specific enough to allow comparisons across participants but also general enough to incentivize further exploration of the tool.
Users completed the tasks only in the \lexplorer system, 
elaborating on how they would approach a task in their respective baseline setup before starting on it in our prototype.
While working on a specific task, they were asked to be vocal about their current state of mind,
especially about what they were looking for both semantically (e.g., a definition) and technically (e.g., a functionality).
Based on their background, participants voice different thoughts and not everyone comments on every aspect of \lexplorer.
Hence, we cannot draw a complete picture of every participant's perception of the implementation of requirements compared to the quantitative setting.

Notably, the user interactions needed to answer questions in the qualitative design would have allowed users to easily answer the questions in the quantitative design as well. 
However, the qualitative design clearly worked, whereas the quantitative design clearly did not. 
Our explanation is that the quantitative design forced users to rationalize the \emph{How}, 
whereas the qualitative design allowed them to focus on the \emph{Why}. 
In a highly semantic professional domain, procedural knowledge is mostly learned implicitly, the latter feels natural while the former feels forced.

\newcommand{\evaluatioParticipant}[1]{$P_{#1}$\xspace}

\section{Prototype Evaluation}\label{sec:prototype-evaluation}

\begin{figure*}[t]
	\begin{subfigure}{\textwidth}
		\raggedleft
		\includegraphics[width=0.8\linewidth]{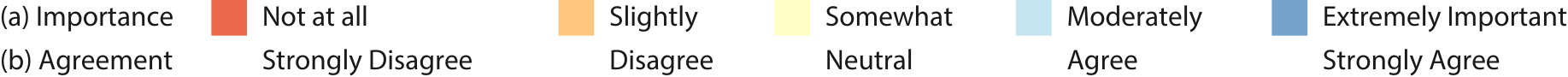}\hspace*{0.75cm}
	\end{subfigure}
	\begin{subfigure}{0.475\textwidth}
		\centering
		\includegraphics[height=3.5cm]{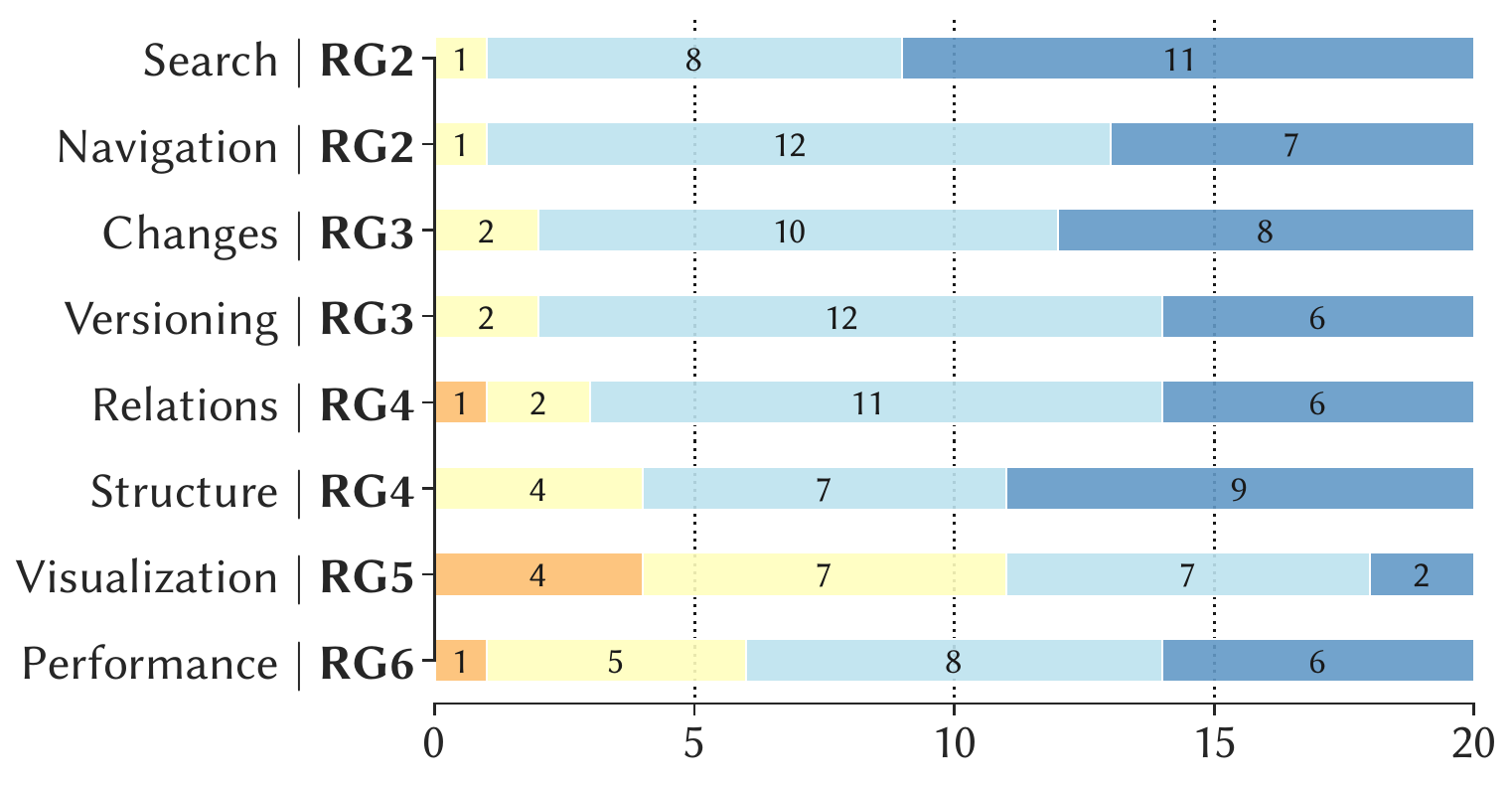}
		\subcaption{Requirements Validation ($n = 20$)}\label{fig:likert-scales:expectations}
	\end{subfigure}
	\begin{subfigure}{0.475\textwidth}
		\centering
		\includegraphics[height=3.5cm]{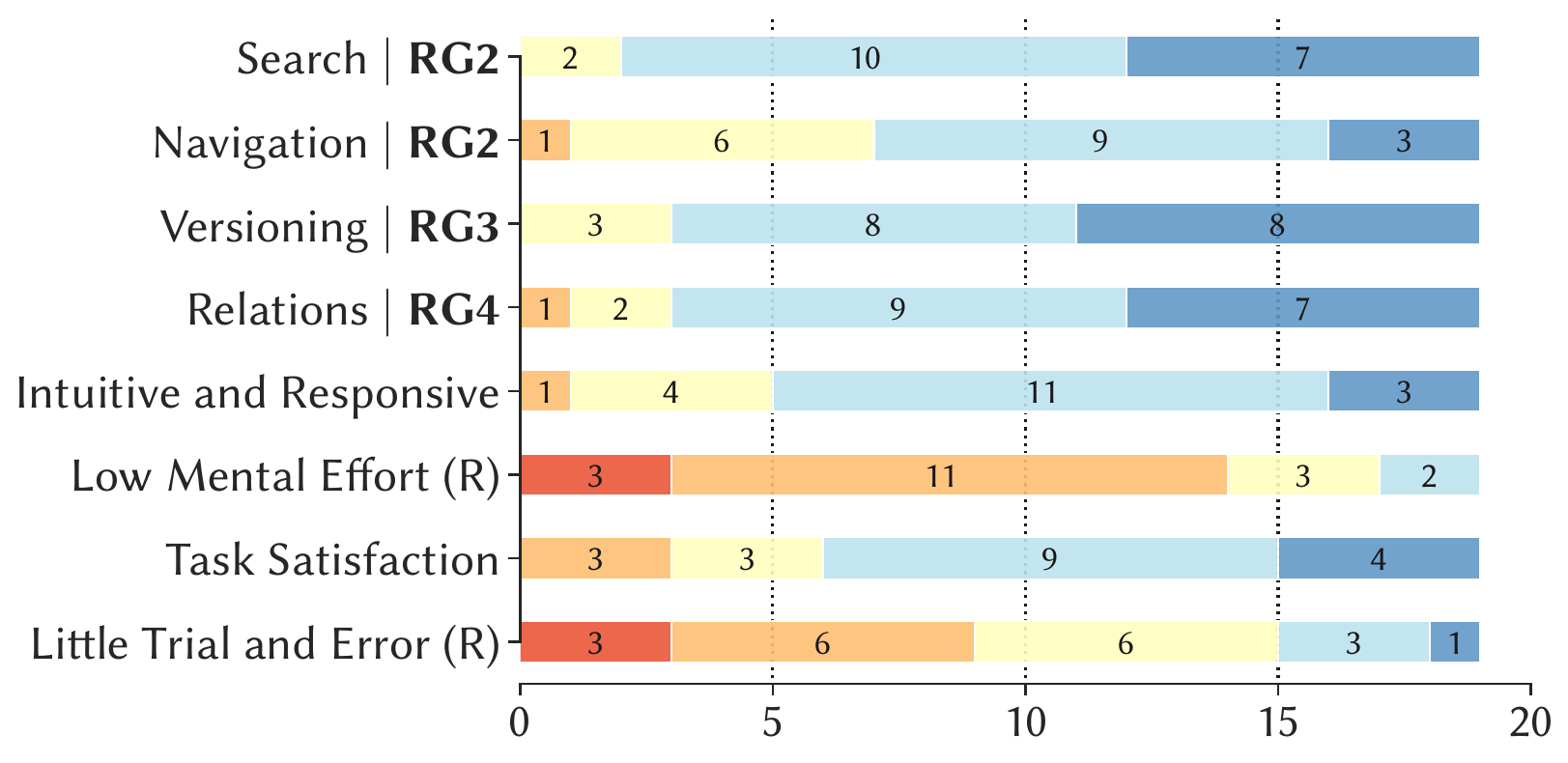}
		\subcaption{Interface Usability ($n = 19$)}\label{fig:likert-scales:feedback}
	\end{subfigure}
	\caption{\textbf{Quantitative results of our prototype evaluation.} Using 5-item Likert scales, (\subref{fig:likert-scales:expectations}) users rated their perceived importance of requirements before conducting the evaluation and (\subref{fig:likert-scales:feedback}) judged the usability of the \lexplorer interface after employing it to work on three legal tasks. One participant did not answer the questionnaire about interface usability. Details on our questionnaires can be found in \Cref{appendix:evaluation-structure}.}\label{fig:likert-scales}
	\Description{The figure shows the quantitative results of the prototype evaluation. Two horizontal bar charts on the left and the right halves of the figure show the exact number of participants and how they voted for each of the eight 5-item Likert scales per bar chart. The left side depicts the requirements validation, the right side shows the interface usability. Above the figure, the legend shows the labels for the 5-item Likert scales, being not at all, slightly, somewhat, moderately, and extremely important for the requirements, as well as strongly disagree, disagree, neutral, agree, and strongly agree for the usability.} %
\end{figure*}

We conducted our prototype evaluation with domain experts ($n = 20$), 
following the qualitative design described in \Cref{sec:evaluation-design}. 
When reporting participants' feedback, 
in the following, 
we write \emph{some} for 2–4 participants, \emph{several} for 5–9 participants, and \emph{many} for 10 or more participants that raised a certain point. 

Four main findings emerged from our evaluation:
First, participants' description of their legal work provided further evidence for the characterization of the legal domain that underlies Adaptive Meaning Construction (AMC).
In particular \emph{normative intertextuality} and \emph{heterogeneous dynamics} were reported by many, while \emph{granular connectivity} was mentioned more selectively across participants. 
Second, many participants responded positively to interface mechanisms that preserved context across documents and exposed fine-grained relationships, especially to reference-navigation features and side-by-side layouts supporting comparisons.
Third, the evaluation revealed some unfamiliarity and tension between visual encoding support for complex relations and the mental effort required to navigate a feature-rich interface, forming two opposing camps among the participants.
Forth, the design process and study further identified five forms of contextual awareness---temporal, procedural, semantic, structural, and provenance awareness---that can guide future interface design for AMC.
While the legal domain was recognized as inherently complex, and several participants sometimes felt overwhelmed with the number of functionalities, they were interested in further engaging with the prototype in the future, and often even requested additional features.
Overall, our evaluation supports the plausibility of our elicited requirements, intent taxonomy, AMC-based design rationale, and initial prototype design, while also revealing limitations and identifying priority areas for future development. 

In the following, we report our participant demographics (\cref{sec:evaluation:setup}),
before elaborating on our findings (\cref{sec:evaluation:results}) with respect to the three evaluation goals formulated in \Cref{sec:evaluation-design}. 
Extended study materials can be found in
\Cref{appendix:participant-demographics-table} and \Cref{appendix:prototype-evaluation}.

\subsection{Demographics and Methodology}\label{sec:evaluation:setup}

Our sample consisted of 16 scholarly, 3 institutional, and one societal actor(s), for a total of $n=20$ participants. 
Scholarly actors were primarily affiliated with 7 different institutions in 2 countries, with core legal education from at least 4 countries, 
including from Northern, Western, and Central Europe. 
More than one third of our experts identified as female~(7/20), the remaining participants identified as male~(13/20). 
At the time of their interview, roughly half of our participants was between 25 and 34 years old~(11/20), the other half between 35 and 84~(9/20).
Fine-grained participant demographics are reported in \Cref{tab:participant-demographics} in \Cref{appendix:participant-demographics-table}.

To target and recruit participants, we followed the same procedure and rationale as for our requirements-elicitation interviews (see~\Cref{sec:requirements:setup}). 
While we deliberately recruited mostly participants who had not participated in our requirements-elicitation interviews, 
we also allowed previous interview participants to evaluate our prototype if they so wished and were available during the evaluation timeframe. 
As a result, one fourth of our final sample (5/20) consisted of participants who had previously participated in our requirements elicitation. 

The \emph{methodology} of our interviews is based on the evaluation design described in \Cref{sec:evaluation-design}.
The majority of interviews lasted around 90 minutes, with the minimum being 45 minutes.
They were exclusively conducted via video call, consensually recorded, and locally transcribed to facilitate further analysis.

Importantly, this prototype evaluation provides a different form of evidence compared to the requirements elicitation in \Cref{sec:requirements}.
While the former captured participants' descriptions of their work and informed the AMC abstraction, the latter challenges participants with these abstractions instantiated as concrete interaction mechanisms while working through realistic legal tasks.
This allowed us not only to examine if the requirements previously identified are aligned with participants' needs, but also to assess where the proposed interaction model supported or conflicted with their reasoning, which design issues emerged in use, and which forms of support were lacking from our prototype.

\subsection{Evaluation Results}\label{sec:evaluation:results}

We group our findings by our three evaluation goals (see~\Cref{sec:evaluation-design}), 
with each group focusing on a different kind of evidence:
For the first goal, \ref{g1}, we draw on what participants reported about their own practice: the tasks they described, the way they reasoned about documents, and the strategies they employ without our prototype.
As this evidence is independent of \lexplorer, we use it to revisit the three properties of legal text that guided our design. 
For the second goal, \ref{g2}, we focus on how participants reacted to the interface design that follows from our intent taxonomy, 
and for the third goal, \ref{g3}, we distill the features requested by participants. 
The codes used in our description (e.g., `B-C' for `uses \emph{Curia} to access European jurisprudence') are resolved to descriptions and participant identifiers in \Cref{tab:evaluation-prototype-results} in \Cref{appendix:prototype-evaluation-results}.

\subsubsection{Evidence for the Validity of Adaptive Meaning Construction~(\ref{g1})}\label{sec:evaluation:results:g1}

Before evaluating the prototype itself, we validated the requirements elicited earlier~(see~\Cref{sec:requirements}) and the domain characterization of law, both of which ground our conception of Adaptive Meaning Construction.

\paragraph{Validation of Requirements}
Participants rated the perceived importance of eight features related to the requirements groups previously identified (\Cref{sec:requirements:findings}) on a 5-item Likert scale from \emph{not at all} to \emph{very much}. 
As illustrated in \Cref{fig:likert-scales:expectations}, 
many participants found requirements pertaining to search and navigation~(\ref{rg:search-and-navigation}), versioning and comparison~(\ref{rg:versioning-and-comparison}), and relationships and context~(\ref{rg:relationships-and-context}) to be of high importance.
For visualization~(\ref{rg:summarization-and-visualization}), expectations were more heterogeneous, with several participants rating the importance as relatively high but several others rating the importance as relatively low.
Many participants also found performance~(\ref{rg:usability-and-performance}) to be of high importance.
Since data access~(\ref{rg:data-access}) is a prerequisite for legal work and several participants have inquired about data completeness and accuracy, we report feedback related to this category separately~(see~\Cref{sec:evaluation:results:g3}).

\paragraph{Validation of Domain Characterization}
During the evaluation, we also elicited participants' descriptions of their past experiences with our baseline system.
Many participants reported to regularly rely on tools beyond EUR-Lex, such as the Google search engine~(C:B-G), Microsoft Word~(C:B-MSWC), and CURIA~(C:B-C).
Their reports allowed us to revisit our assumptions about \emph{heterogeneous dynamics}~(\ref{p1}), \emph{granular connectivity}~(\ref{p2}), and \emph{normative intertextuality}~(\ref{p3}) in law.
Many participants showed an intuition for \emph{heterogeneous dynamics} as they referred to situations in which they had to directly compare two versions of consolidated documents to detect legal changes~(C:MD-AC, C:MD-UCRC, and C:MD-UCT).
In particular, they described \emph{capturing and communicating change} as the core of doctrinal research~(C:MD-AC). 
Some participants specifically articulated their gratitude for the fine-grained resolution of references~(C:R-FG) needed to accommodate \emph{granular connectivity}. 
Many participants also mentioned references as useful for research in case law~(C:R-CL), and several reported working closely with references in their baseline to uncover relations between documents~(C:R-MP, C:R-OT, and C:R-TS), 
indicating the relevance of reference structure in the context of \emph{normative intertextuality}. 
Beyond references, intertextuality also manifests in the desire to consult administrative documents~(C:D-DA) and in the comparison of meaning differences across language versions of EU legal documents~(C:MD-UCLANG).

\subsubsection{Usability, Effectiveness, and Feedback on Design Choices in \lexplorer (\ref{g2})}\label{sec:evaluation:results:g2}

Beyond the usefulness of the reference functionality, we also collected participants' perspectives on the overall usability of \lexplorer.

\paragraph{Perceived Usability and Effectiveness of the Interface}
After concluding the tasks, participants were able to judge the implementation of requirements and the perceived usability by rating their agreement with eight statements on a 5-item Likert scale from \emph{strongly disagree} to \emph{strongly agree}~(see~\Cref{fig:likert-scales:feedback}).
Many participants \emph{(strongly) agreed} that search~(\ref{rg:search-and-navigation}), navigation~(\ref{rg:search-and-navigation}), versioning~(\ref{rg:versioning-and-comparison}), and relation~(\ref{rg:relationships-and-context}) features were implemented in a usable manner, while the remaining users were either \emph{neutral} in their agreement or \emph{disagreed}.
About three quarters of the participants~(14/19) also \emph{(strongly) agreed} that the interface of \lexplorer is intuitive and responsive, while only one person \emph{disagreed} and four were \emph{neutral}.
Many participants also expressed satisfaction with their task performance.
However, many participants \emph{(strongly) disagreed} with the tasks requiring little mental effort and several admitted to a \emph{trial-and-error} strategy when navigating the prototype.

\paragraph{Reception of Design Choices}
Many participants appreciated the functionalities regarding \emph{references}~(C:R-AR, C:R-FG, and C:R-U), with several engaging intuitively with them~(C:R-IT and C:R-R).
While participants often manually parse references when using the baseline~(C:R-MP and C:R-TS), \lexplorer resolves references to fine-grained structural entities, which some participants appreciated~(C:R-FG). 
This even caused some participants to feel that the way references are implemented in our prototype increases their productivity~(C:R-IP).

In addition to the references, the \emph{side-by-side layout} supporting comparison across versions and documents in the one- and few-documents views~(C:MD-SBS), was described positively by many participants.
Several of them also found the temporal comparison to be intuitive~(C:MD-ITC), constituting an update over the alternative strategy of opening multiple tabs in their browser~(C:B-BT) or comparing versions using tools like Microsoft Word~(C:B-MSWC).
To make sense of the references and contents within the side-by-side layout, some participants appreciated the \emph{color coding} of document types~(C:ST-CC) and several others appreciated the interactive \emph{table of contents}~(C:O-A).
The visual representation of the table of contents in the form of a glyph enhanced with document-internal relationships was also found useful~(C:V-UI) and easy to navigate~(C:V-NSV) by several participants, while others reported little imagination for \emph{visualization}~(C:V-LI) or a harder time getting along with it~(C:V-NUI and C:V-SV).
\paragraph{Tension between Expressiveness and Complexity}
The vast number of functionalities contributed to a feeling of \emph{overwhelm} in several participants~(C:C-O), while individuals also appreciated the feature richness~(C:C-AC) of the interface.
When asked about why they felt overwhelmed, several participants referred to law being inherently complex~(C:C-LIC) and showed confidence that getting more familiar with the prototype was only a matter of time~(C:C-MOT).
While this qualitative insight does not rule out room for improvement on the prototype's usability and intuitiveness, it suggests that there is an inherent trade-off between feature richness and mental load.

Taken together, our evaluation provides evidence that the intents and properties underlying AMC correspond to recurring patterns in legal work, while demonstrating that contextualized cross-document interaction can support these patterns in practice.
While our study exposes a field of tension between expressive support and interaction complexity, it also identifies five forms of awareness that are promising for future research on AMC, delineated below.
These findings inform both the continued development of \lexplorer and the broader discussion of interface design in \Cref{sec:discussion}.

\subsubsection{Beyond the Prototype: Five Forms of Awareness Required for AMC (\ref{g3})}\label{sec:evaluation:results:g3} %

Our findings indicate that five forms of awareness are required for AMC, each constituting an important direction for the future development of \lexplorer.

\paragraph{Temporal Awareness}
Several participants expressed an interest in tracking individual provisions or legal concepts over \emph{longer} periods of time (C:MD-TF) to trace the evolution of legislation generally or individual documents specifically. 
Some participants suggested timelines as visual components to navigate such developments (C:MD-TL).
Future versions of \lexplorer should therefore investigate how temporal aspects can be sufficiently captured (given the limited amount of public intermediate versions or structured metadata available) and how a temporal view could expose the trajectory visually.
For example, one could imagine a timeline combining version histories with the surrounding document relationships, enabling the investigating not only \emph{what} changed, but also how changes relate over specific time-frames.

\paragraph{Procedural Awareness} 
Beyond temporal awareness, \lexplorer could benefit from embracing that legal documents hardly ever exist in isolation: legislative, judicial, and administrative procedures come in \emph{document families}, i.e., sets of procedurally related documents, that are inherently and meaningfully interconnected, even if they are not explicitly linked via cross-references.
The demand to include such documents was reflected in participants' questions,
some asking explicitly for procedural documents (C:D-DP), 
others inquiring about administrative documents (C:D-DA). 
Future work could therefore investigate how sequences and roles of documents could be effectively represented and exposed. 
This would enhance the current reference-centered representation with procedural context, allowing users to navigate the genesis of legal documents, rather than only its explicit connections \emph{after} publication (e.g., associations' statements that influence the legislative process or drafts that are traces of pre-publication development).

\paragraph{Semantic Awareness}
A third opportunity for improvement concerns the \emph{heterogeneity} of legal documents themselves.
Our current approach provides a shared interaction vocabulary across document types. 
However, legislation, jurisprudence, and other legal materials can have different semantic roles and expose different kinds of relevant metadata and relationships.
Future iterations could therefore explore how views and interactions can be specifically tied to the document under consideration, while simultaneously preserving the coordinated interaction model across our presented views.
Several participants showed intuitive awareness of document-types (C:ST-DT),
some specifically suggested including metadata relevant to certain kinds of documents (C:ST-DM).
The benefit of semantic awareness is not limited to the presentation of individual documents,
but it extends to their localization in the broader legal context.
Regulatory frameworks often cross jurisdictional borders---%
for example when EU law interacts with national implementations or when international law provides a further source of legislative material.
\evaluatioParticipant{9} revealed that they almost never had to consider \emph{only} EU law.
Therefore, developing a document-type- and legal-system-aware interaction model would be an important step toward supporting genuinely heterogeneous legal document landscapes.
A platform like \lexplorer with access to \emph{national}, \emph{international}, and \emph{regional} law
was predicted to be a `gamechanger'  (\evaluatioParticipant{18})
that would cause `storms of jubilation' (\evaluatioParticipant{5});
\evaluatioParticipant{29}~called the prospect `utopically beautiful'.

\paragraph{Structural Awareness}
\lexplorer presents documents and parses references
at their most granular level,
i.e., the lowest possible
sub-section (in legislation)
or
a specific paragraph (in court decisions).
Participant comments to further develop this structural path
can be summarized as requesting \emph{more exploratory control over structural information}.
With regard to incoming references,
for example,
participants wanted to search for a specific keyword within the references (C:R-WTS)
or apply custom ordering to the list of references (C:R-WOR).
Some participants suggested extending
structural paths
by including \emph{reference chains}, 
which would avoid multiple (manual) hops (C:R-WRC).
Future work should therefore explore how to harness
the users' semantic sense of orientation
to even better exploit \emph{structural} insights,
e.g., routes through document landscapes.

\paragraph{Provenance Awareness}
\lexplorer currently preserves context between coordinated views and allows users to collect documents in dossiers, but it provides limited support for externalizing how a user reaches a conclusion.
Future systems could more explicitly track and present user navigation paths
and potentially
incorporate a graph-based, non-linear history.
Leveraging \emph{analytical provenance}~\cite{northAnalyticProvenanceProcess+interaction+insight2011}, a system can trace semantically-informed, behavior-induced interpretive paths through documents and relationships, capturing a particular legal interpretation, helping users return to previous legal hypotheses, communicate an argument to collaborators, or provide guided entry points.

	\section{Discussion}\label{sec:discussion}

While our research and design process allowed deep insights into the needs of our target user group and yielded a prototype appreciated by that constituency (\Cref{sec:prototype-evaluation}), 
our study has several limitations that also highlight directions for future work (\Cref{sec:discussion:limitations}). 
Beyond the specific limitations of our study, 
some valuable learnings concern the work process itself (\Cref{sec:discussion:challenges}). 

\subsection{Limitations and Future Work}\label{sec:discussion:limitations}
As the limitations of our prototype implementation and related next steps have already been presented in the context of our evaluation (\Cref{sec:prototype-evaluation}),  
the following discussion focuses on three groups of limitations and potentials:
intent abstraction, data coverage, and user-study methodology.

\subsubsection{Intent Abstraction}\label{sec:discussion:limitations:interaction-model}

One limitation concerns the scope and abstraction of our AMC model.
Adaptive Meaning Construction is primarily \textbf{derived from elicited needs} in legal work (see \Cref{sec:requirements}) and therefore reflects existing practices and struggles of our participants. 
While this grounding was intentional, it may also privilege interaction patterns that users already know over less familiar alternatives.
Furthermore, mapping heterogeneous legal workflows onto one unified intent abstraction necessarily involves prioritization. 
In particular, the \textbf{focus} of our abstraction lies on information discovery and scoping as well as analysis and interpretation, while synthesis, documentation, and later stages of legal reasoning remain largely outside its scope.
A related limitation concerns the generalizability to different domains. Although we frame AMC as a broader problem class, our concrete formulation and implementation are grounded in EU legal work.
Thus, while we expect parts of the model to transfer to other legal systems (such as those typically characterized as following the common-law tradition, e.g., the United States) and other domains in which meaning must be constructed across interdependent documents, 
this has not yet been demonstrated empirically.
Therefore, exploring the \textbf{generalizability of AMC} to other domains constitutes a fruitful avenue for further research.

\subsubsection{Data Coverage}\label{sec:discussion:limitations:data-coverage}
Another limitation concerns the \textbf{availability}, completeness, and quality of the legal \textbf{data} underlying our prototype.
Many participants emphasized that their trust in a legal information system depends on its ability to provide information that is comprehensive and correct.
While \lexplorer currently focuses on EU legal sources, legal work frequently combines EU, national, regional, and international law, as well as scholarly, commercial, and other material.
More generally, access to sufficiently complete legal data is itself often restricted: 
In many jurisdictions, even elementary legal materials (such as all consolidated versions of a legislative text) are only available through commercial providers (e.g., juris and beck-online in Germany, Westlaw and LexisNexis in the US, or Wolters Kluwer), 
and many court decisions remain unpublished due to anonymization concerns \cite{terzidou2023automated,coupette2018quantitative}.
As a result, while the practical usefulness of our approach depends heavily on sufficient access to (open) data, 
such access is currently mostly restricted to commercial platforms and licensing models.

Beyond basic data access, several of our interactions rely on \textbf{structured information} that is still not readily available from primary sources. 
Hence, our approach currently requires heavy pre-processing in our backend---%
from parsing documents in different formats into a common schema 
to extracting fine-grained references, document-type-specific metadata, and semantic information---, 
which introduces additional dependencies regarding correctness, maintenance, and scalability. 
This can complicate the evaluation of our interface, 
as incomplete data can cause a user to reject an otherwise useful workflow, 
while carefully curated prototype data may overstate its usefulness under production conditions.
Data coverage also affects the validity of our \textbf{baseline comparison}.
We selected EUR-Lex because it is mostly available as open data, yet participants described heterogeneous baseline setups, where they manually combine legal information systems, search engines, online browsing, and text-editing tools, yielding an information advantage.

As the usefulness of any prototype depends substantially on the completeness, structure, and breadth of the underlying data, 
a high-impact research direction will investigate how to move toward a \textbf{more comprehensive coverage of EU legal material}, 
including administrative and procedural documents and intermediate versions as well as structured metadata.
Beyond EU-level sources, future versions of \lexplorer could incorporate other \textbf{national and international legal documents}.
This is particularly important for cases where jurisdiction is not limited to a single country or legal system.
Extending \lexplorer to other legal systems would also offer the opportunity to validate empirically its usefulness beyond the continental legal tradition.

\subsubsection{User-Study Methodology}\label{sec:discussion:evaluation}

Our user-study methodology has several limitations, 
most of which are by design. 

\paragraph{Predominantly Qualitative Evaluation}
One major limitation is that our \textbf{prototype evaluation} is predominantly \textbf{qualitative}.
This decision is informed by our trials in the evaluation-design phase of our prototype evaluation (\Cref{sec:evaluation-design}): 
forcing a quantitative evaluation with elements such as task answers, completion times, click streams, and measures of correctness and completeness, would have sacrificed ecological validity.
Beyond substantive concerns, 
another major obstacle in performing a quantitative evaluation was the participants' extremely \textbf{heterogeneous baseline setup} (see also \Cref{sec:requirements}). 
This made accurately measuring and comparing baselines (both across participants and to \lexplorer) rather challenging.
Even when setting a common baseline, such as restricting the workflow exclusively to EUR-Lex, 
capturing granular quantitative information from that baseline for remote participants worldwide would have been unrealistic---%
first due to technical limitations of the baseline platform, 
and second because forcing participants to abandon their regular work environments would have yielded an unfair comparison. 
Hence, we ultimately decided on a qualitative evaluation, both substantive and technical reasons.
As a necessary consequence, our results mainly support claims about perceived usefulness, interaction fit, and observed behavior, 
rather than about statistically robust improvements in efficiency or result quality.

\paragraph{Ecological Validity}
Participants worked on \textbf{researcher-defined tasks} covering several areas of EU law, rather than bringing their own real-world problems.
While our tasks were designed by the legal scholars in our team to reflect recurring legal activities, 
and the common-task setup was chosen to enhance cross-participant comparability (see \Cref{sec:evaluation-design}), 
they cannot fully reproduce the open-endedness, interruptions, and uncertainty of everyday legal work.
Similarly, the think-aloud setting and the \textbf{guided introduction} to the systems may have influenced interaction behavior in a systematic manner, 
and our short evaluation phases cannot capture \textbf{longer-term adaptation} when tools are integrated into established workflows.
Also, our evaluation cannot reliably determine how much of the observed mental effort resulted from the domain, an unfamiliarity with the prototype, or avoidable interaction-design complexity; disentangling these factors would also require longitudinal evaluations.

\paragraph{Sampling Bias}
Both of our samples were \textbf{skewed} toward \textbf{scholarly} participants. 
While legal scholars are the primary target group of our prototype, 
we hope that our work will also support institutional and civil-society actors, 
which were underrepresented in our samples. 
Furthermore, our samples exhibit skew toward \textbf{male} participants, 
underrepresenting individuals identifying with other genders. 
While this reflects a skew in the underlying target population, 
we acknowledge that a different gender balance may have yielded different results. 
Since the focus of our prototype was EU law, our samples are also tilted toward \textbf{WEIRD} demographics. 
\textbf{Network-based recruitment} may additionally favor participants already interested in legal technology, 
while prior visualization or computational experience may influence how easily participants engage with the interaction techniques.
By design, a subset of participants also took part in both of our studies, which potentially introduces subtle confirmation effects for those individuals.
Finally, \textbf{novelty and learning effects} are difficult to disentangle from the properties of the interface itself.
Several interactions differed substantially from participants' established practices.
Some participants initially struggled with diff views or visual navigation, although many expected these difficulties to diminish with greater familiarity.
At the same time, novel functionality such as integrated comparison or cross-reference navigation may also have contributed to particularly positive first impressions.
Our study therefore cannot determine whether these reactions represent long-term improvements in usability or short-term learning and novelty effects, 
nor can it establish whether initial acceptance translates into sustained adoption or workflow change. 
Addressing these questions will require longer-term observation and continued use embedded in participants' daily work.

\paragraph{Future Work}
Future evaluation endeavors can leverage participatory approaches and longitudinal studies.
We imagine \textbf{multi-day workshops or hackathons} with legal scholars as a useful first step for collaboratively developing our approach further.
This has the potential to benefit and evaluate novel concepts---such as \emph{semantic tours} or \emph{document-type-adaptive views}---that go beyond interaction patterns already familiar to participants.
Moreover, a long-term study of the prototype when deployed as an \textbf{embedded beta system} that lawyers can use in their own ongoing work will provide additional valuable insights.
This would expose \lexplorer to the open-endedness, interruptions, heterogeneous tool usage, and evolving questions of actual legal research, 
addressing the limitations of our present evaluation.
\textbf{Longitudinal observation} could investigate whether some of the observed initial difficulties with unfamiliar interaction techniques diminish with experience, whether positive first impressions translate into sustained adoption, and how users incorporate the system into their existing  ecosystems.

\subsection{Challenges Encountered and Lessons Learned}\label{sec:discussion:challenges}

Having addressed the limitations of our study and the resulting promising directions for future work, 
we now turn to our \textbf{meta learnings}. 
A first meta learning was already discussed in \Cref{sec:evaluation-design,sec:discussion:evaluation}: 
\emph{When evaluating in a highly semantic domain, resist quantification.}
In the following, 
we discuss two additional key challenges we encountered alongside the strategies we used to overcome them. 

\subsubsection{Negotiating Methodological and Epistemic Authority: Devise Middle-Layer Abstractions}

Working across disciplinary boundaries naturally comes with \textbf{communication challenges} \cite{monteiro2009managing}. 
In our project, we learned that these challenges persist \emph{even if} one side knows how to speak the \emph{language} but not the \emph{specialization} of the other side. 
(In our case, the legal scholars involved in the project had a high level of familiarity with different areas in computer science but no expertise in visual analytics.)
We believe the reasons for this to be twofold. 
First, one needs to know not only \emph{how} to translate between domains but also \emph{when} to translate, 
which is easier said than done. 
Second, and more importantly, 
in a transdisciplinary collaboration aiming to leverage the expertise of one discipline to address problems faced by the other discipline, 
methodological and epistemic authority need to be constantly (re)calibrated---%
i.e., there are not just communication challenges but \textbf{power struggles}. 
For example, 
when the lawyers on the team had a strong intuition that a quantitative evaluation of \lexplorer would not work, 
they initially yielded to the visual-analytics experts, 
assuming that they would `know better' by the nature of their expertise, 
and accepting the assertion that `this is how you evaluate things' in interface design. 
On a more positive note, the features of our prototype that were most appreciated in our evaluation (e.g., the ribbons to access incoming references) turned out to be exactly those that were subject to the most detailed discussions across the disciplinary divide---%
because these features ended up reflecting a true \textbf{alignment of mental models}. 
In our case, the path toward mental-model alignment was rather convoluted, 
and it led us to AMC as a tool to help us `think across the aisle'. 
Adaptive Meaning Construction is a \textbf{middle-layer abstraction}, 
i.e., a domain- and method-agnostic concept capturing those parts of domains and methods that are required to bridge the gap between domain-specific and methodological expertise. 
Our experience suggests that middle-layer abstractions hold significant potential to facilitate cross-disciplinary collaboration, in interface design and beyond. 

\subsubsection{Building Interfaces to Push Domain Boundaries: Internalize Interdisciplinarity}

Continuing our reflection on the relationship between expertise and (research) authority, 
another challenge we faced was \textbf{deference to the domain}. 
The visual-analytics researchers on the team approached the domain with the goal of supporting how experts currently \emph{do work}, 
rather than how they \emph{should be working}. 
While this approach is inherently prone to \textbf{XY problems},
in our case, it also led to conflicts with the legal scholars, 
who were determined to push the methodological boundaries of their field. 
While we eventually settled for pushing the boundaries at least a little bit, 
the situation we encountered reflects a broader problem in the design of interfaces for domain experts: 
By seeking to help experts work better, 
interface designers tend to reinforce the methodological \emph{status quo} in the domain, promoting \textbf{methodological conservatism}. 
One way out of this conundrum is to \textbf{internalize interdisciplinarity} by involving researchers who are at the methodological frontier of their domain in the tool-development process. 
However, this immediately clashes with the established \textbf{culture of neutrality} in technical domains: 
By which standards should we judge a prototype when part of the team is among the intended users? 
How the scientific community answers this question will impact the incentives to engage in truly interdisciplinary work.

	\section{Conclusion}\label{sec:conclusion}

Legal information systems have traditionally been designed under the assumption that users arrive with an information need that can be progressively resolved through search and retrieval, which constitute the most deployed features of such systems.
Our findings, based on the literature and interviews ($n=15$) with scholars, challenge this assumption for legal research:
Legal scholars often start with only a partial understanding of what is relevant, and that understanding evolves when reading through documents, their versions, following references, and contemplating legal concepts.
Therefore, legal work often involves not just locating but continuously interpreting and reorganizing information.
We describe this class of activity as Adaptive Meaning Construction (AMC), 
distinguishing law from other text-heavy domains via its heterogeneous dynamics, granular connectivity, and normative intertextuality.
In doing so, we shift our design problem from supporting the reading of legal documents toward supporting the evolving analytical intents through which legal meaning is constructed.

Our \lexplorer system demonstrates how this shift can inform interface design.
Rather than organizing interaction around isolated search results and individual documents, our system combines exploration, close reading, comparison, and relationship navigation across coordinated one-, few-, and many-document views of legal documents, seamlessly transitioning between text- and data-focused modes.
Our evaluation ($n=20$) indicates that interactions centered on cross-references, contextual navigation, and modern forms of fine-grained version comparison correspond closely to practices already embedded in legal analysis.
At the same time, it shows that structural information becomes useful only when its presentation remains connected to the semantic questions pursued by the interface user.  
Supporting complex analytical work therefore requires more than exposing additional data or relationships. 
Instead, an interface must make those structures \emph{actionable} within the user's process of meaning construction.

The perspective we develop has implications beyond the specific design of \lexplorer.
A central challenge we encountered was that expert practices are difficult to translate directly into interface requirements.
Procedural knowledge in legal scholarship is often tacit and implicit, and attempts to reduce that work to predefined, concretely measurable tasks risk stripping away crucial interpretive processes, biasing results.
Our findings suggest that the practice of interface design for such domains benefits from an intermediate level of abstraction:
It has to be sufficiently grounded in domain practice to preserve meaningful distinctions, 
yet sufficiently independent of domain-specific terminology to guide interaction design and enable transfer.
AMC represents one such abstraction, while our design process illustrates the methodological work required to construct it.

Our prototype \lexplorer remains an initial step toward a richer environment for legal analysis.
As the volume of legal data grows and legal text becomes more interconnected, the central challenge for interactive systems is increasingly not how to return more `relevant' documents, but how to help users understand what those documents mean in relation to one another, 
supporting them along the way.
For legal scholarship, and potentially for other forms of expert analytical work, this suggests a broader shift in interface design:
from systems that primarily retrieve information toward systems that actively support the construction of meaning.

	\section*{Acknowledgments}\label{sec:acknowledgments}

The authors gratefully acknowledge the research assistance of Julia Görlach.
This work has been partly funded by the Federal Ministry of Research, Technology and Space (BMFTR) in MATRIX-PRO (FKZ: 13N17718) and under Germany's Excellence Strategy - EXC 2117 - 422037984. 
This work is supported by ERC grant \textsc{CompLex} (Grant Number: \href{https://cordis.europa.eu/project/id/101221337}{101221337}).
Funded by the European Union. Views and opinions expressed are however those of the author(s) only and do not necessarily reflect those of the European Union or the European Research Council Executive Agency. Neither the European Union nor the granting authority can be held responsible for them.

	\bibliographystyle{ACM-Reference-Format}
	\bibliography{references}	

@article{sedlmair2012design,
	title = {Design {Study} {Methodology}: {Reflections} from the {Trenches} and the {Stacks}},
	volume = {18},
	copyright = {https://ieeexplore.ieee.org/Xplorehelp/downloads/license-information/IEEE.html},
	issn = {1077-2626},
	shorttitle = {Design {Study} {Methodology}},
	url = {http://ieeexplore.ieee.org/document/6327248/},
	doi = {10.1109/TVCG.2012.213},
	number = {12},
	urldate = {2025-10-30},
	journal = {IEEE Transactions on Visualization and Computer Graphics},
	author = {Sedlmair, Michael and Meyer, Miriah and Munzner, Tamara},
	year = {2012},
	pages = {2431--2440},
}

@article{furst2025challenges,
	title = {Challenges and {Opportunities} for {Visual} {Analytics} in {Jurisprudence}},
	issn = {0924-8463, 1572-8382},
	url = {https://link.springer.com/10.1007/s10506-025-09494-2},
	doi = {10.1007/s10506-025-09494-2},
	language = {en},
	urldate = {2025-11-26},
	journal = {Artificial Intelligence and Law},
	author = {Fürst, Daniel and El-Assady, Mennatallah and Keim, Daniel A. and Fischer, Maximilian T.},
	year = {2025},
}

@article{katz2020complex,
	title    = {{Complex Societies and the Growth of the Law}},
	volume   = {10},
	issn     = {2045-2322},
	doi      = {10.1038/s41598-020-73623-x},
	language = {en},
	number   = {1},
	journal  = {Scientific Reports},
	author   = {Katz, Daniel Martin and Coupette, Corinna and Beckedorf, Janis and Hartung, Dirk},
	year     = {2020},
	pages    = {18737}
}

@article{coupette2021measuring,
	title      = {Measuring {Law} {Over} {Time}: {A} {Network} {Analytical} {Framework} with an {Application} to {Statutes} and {Regulations} in the {United} {States} and {Germany}},
	volume     = {9},
	issn       = {2296-424X},
	shorttitle = {Measuring {Law} {Over} {Time}},
	url        = {https://www.frontiersin.org/articles/10.3389/fphy.2021.658463/full},
	doi        = {10.3389/fphy.2021.658463},
	journal    = {Frontiers in Physics},
	author     = {Coupette, Corinna and Beckedorf, Janis and Hartung, Dirk and Bommarito, Michael and Katz, Daniel Martin},
	year       = {2021},
	pages      = {658463}
}

@article{punder2026power,
	title={The Power of Network Pluralism: Multi-Perspective Modeling of Heterogeneous Legal Document Networks},
	author={P{\"u}nder, Titus and Coupette, Corinna},
	volume = {99},
	ISSN = {1434-6036},
	DOI = {10.1140/epjb/s10051-026-01178-3},
	number = {9},
	journal = {The European Physical Journal B},
	publisher = {Springer Science and Business Media LLC},
	year={2026}
}

@article{coupette2023law,
	title={Law Smells: Defining and Detecting Problematic Patterns in Legal Drafting},
	author={Coupette, Corinna and Hartung, Dirk and Beckedorf, Janis and B{\"o}ther, Maximilian and Katz, Daniel Martin},
	journal={Artificial Intelligence and Law},
	volume={31},
	number={2},
	pages={335--368},
	year={2023},
	publisher={Springer}
}

@inproceedings{pirolli2005sensemaking,
  title = {The Sensemaking Process and Leverage Points for Analyst Technology as Identified through Cognitive Task Analysis},
  booktitle = {Proceedings of International Conference on Intelligence Analysis},
  author = {Pirolli, Peter and Card, Stuart},
  volume = {5},
  pages = {2--4},
  year = {2005}
}

@techreport{cook2005illuminating,
  title = {Illuminating the {{Path}}: {{The Research}} and {{Development Agenda}} for {{Visual Analytics}}},
  author = {Cook, Kristin A and Thomas, James J},
  institution = {Pacific Northwest National Lab (PNNL), Richland, WA, United States},
  year = {2005}
}

@article{kaixuAnalyticProvenanceSensemaking2015,
  title = {Analytic {{Provenance}} for {{Sensemaking}}: {{A Research Agenda}}},
  author = {{Kai Xu} and Xu, Kai and {Simon Attfield} and Attfield, Simon and {T. J. Jankun-Kelly} and Jankun-Kelly, T. J. and {Ashley Wheat} and Wheat, Ashley and {Phong Q. Nguy\^e\~n} and Nguyen, Phong H. and {Nallini Selvaraj} and Selvaraj, Nallini},
  volume = {35},
  pages = {56--64},
  doi = {10.1109/mcg.2015.50},
  year = {2015},
  journal = {IEEE Computer Graphics and Applications}
}

@article{lettieriLegalMacroscopeExperimenting2017,
  title = {The Legal Macroscope: {{Experimenting}} with Visual Legal Analytics},
  author = {Lettieri, Nicola and Altamura, Antonio and Malandrino, Delfina},
  shortjournal = {Information Visualization},
  volume = {16},
  pages = {332--345},
  doi = {10.1177/1473871616681374},
  year = {2017},
  journal = {Information Visualization}
}

@article{kleinMakingSenseSensemaking2006a,
  title = {Making {{Sense}} of {{Sensemaking}} 2: {{A Macrocognitive Model}}},
  author = {Klein, G. and Moon, B. and Hoffman, R.R.},
  shortjournal = {IEEE Intell. Syst.},
  volume = {21},
  pages = {88--92},
  publisher = {{Institute of Electrical and Electronics Engineers (IEEE)}},
  doi = {10.1109/mis.2006.100},
  year = {2006},
  journal = {IEEE Intelligent Systems}
}

@article{resckLegalVisExploringInferring2023,
  title = {{{LegalVis}}: {{Exploring}} and {{Inferring Precedent Citations}} in {{Legal Documents}}},
  author = {Resck, Lucas E. and Ponciano, Jean R. and Nonato, Luis Gustavo and Poco, Jorge},
  shortjournal = {IEEE Trans. Visual. Comput. Graphics},
  volume = {29},
  pages = {3105--3120},
  doi = {10.1109/TVCG.2022.3152450},
  year = {2023},
  journal = {IEEE Transactions on Visualization and Computer Graphics}
}

@inproceedings{shrinivasanSupportingAnalyticalReasoning2008,
  title = {Supporting the Analytical Reasoning Process in Information Visualization},
  booktitle = {Proceedings of the {{SIGCHI Conference}} on {{Human Factors}} in {{Computing Systems}}},
  author = {Shrinivasan, Yedendra Babu and Van Wijk, Jarke J.},
  date = {2008-04-06},
  pages = {1237--1246},
  publisher = {ACM},
  doi = {10.1145/1357054.1357247},
  year = {2008},
  address = {Florence Italy}
}

@article{batesDesignBrowsingBerrypicking1989,
  title = {The Design of Browsing and Berrypicking Techniques for the Online Search Interface},
  author = {Bates, Marcia J.},
  shortjournal = {Online Review},
  volume = {13},
  pages = {407--424},
  doi = {10.1108/eb024320},
  year = {1989},
  journal = {Online Review}
}

@inproceedings{lacavaLawNetVizWebbasedSystem2022,
  title = {{{LawNet-Viz}}: {{A Web-based System}} to {{Visually Explore Networks}} of {{Law Article References}}},
  booktitle = {Proceedings of the 45th {{International ACM SIGIR Conference}} on {{Research}} and {{Development}} in {{Information Retrieval}}},
  author = {La Cava, Lucio and Simeri, Andrea and Tagarelli, Andrea},
  pages = {3300--3305},
  publisher = {ACM},
  doi = {10.1145/3477495.3531668},
  year = {2022}
}

@inproceedings{duckFindingNeedlesDocument2025,
  title = {Finding {{Needles}} in {{Document Haystacks}}: {{Augmenting Serendipitous Claim Retrieval Workflows}}},
  booktitle = {Proceedings of the 2025 {{CHI Conference}} on {{Human Factors}} in {{Computing Systems}}},
  author = {D\"uck, Moritz and Holter, Steffen and Chan, Robin Shing Moon and Sevastjanova, Rita and El-Assady, Mennatallah},
  pages = {1--17},
  publisher = {ACM},
  doi = {10.1145/3706598.3713715},
  year = {2025},
}

@inproceedings{fokQlarifyRecursivelyExpandable2024,
  title = {Qlarify: {{Recursively Expandable Abstracts}} for {{Dynamic Information Retrieval}} over {{Scientific Papers}}},
  booktitle = {Proceedings of the 37th {{Annual ACM Symposium}} on {{User Interface Software}} and {{Technology}}},
  author = {Fok, Raymond and Chang, Joseph Chee and August, Tal and Zhang, Amy X. and Weld, Daniel S.},
  pages = {1--21},
  publisher = {ACM},
  doi = {10.1145/3654777.3676397},
  year = {2024},
}

@inproceedings{kimDataDiveSupportingReaders2024,
  title = {{{DataDive}}: {{Supporting Readers}}' {{Contextualization}} of {{Statistical Statements}} with {{Data Exploration}}},
  booktitle = {Proceedings of the 29th {{International Conference}} on {{Intelligent User Interfaces}}},
  author = {Kim, Hyunwoo and Le, Khanh Duy and Lim, Gionnieve and Kim, Dae Hyun and Hong, Yoo Jin and Kim, Juho},
  pages = {623--639},
  publisher = {ACM},
  doi = {10.1145/3640543.3645155},
  year = {2024},
}

@inproceedings{choeCrossLitConnectingVisual2026,
  title = {{CrossLit}: Connecting Visual and Textual Sensemaking for Literature Review},
  booktitle = {Proceedings of the 2026 {{CHI Conference}} on {{Human Factors}} in {{Computing Systems}}},
  author = {Choe, Kiroong and Kim, Eunhye and Kim, Min Hyeong and Hwang, Suyeon and Park, Sangwon and Kim, Nam Wook and Seo, Jinwook},
  pages = {1--30},
  publisher = {ACM},
  doi = {10.1145/3772318.3791418},
  year = {2026},
}

@inproceedings{kangSynergiMixedInitiativeSystem2023,
  title = {Synergi: {{A Mixed-Initiative System}} for {{Scholarly Synthesis}} and {{Sensemaking}}},
  booktitle = {Proceedings of the 36th {{Annual ACM Symposium}} on {{User Interface Software}} and {{Technology}}},
  author = {Kang, Hyeonsu B and Wu, Tongshuang and Chang, Joseph Chee and Kittur, Aniket},
  pages = {1--19},
  publisher = {ACM},
  doi = {10.1145/3586183.3606759},
  year = {2023},
}

@inproceedings{horvathUsingAnnotationsSensemaking2022,
  title = {Using {{Annotations}} for {{Sensemaking About Code}}},
  booktitle = {Proceedings of the 35th {{Annual ACM Symposium}} on {{User Interface Software}} and {{Technology}}},
  author = {Horvath, Amber and Myers, Brad and Macvean, Andrew and Rahman, Imtiaz},
  pages = {1--16},
  publisher = {ACM},
  doi = {10.1145/3526113.3545667},
  year = {2022},
}

@article{mclachlanVisualisationLawLegal2021,
  title = {Visualisation of Law and Legal {{Process}}: {{An}} Opportunity Missed},
  author = {McLachlan, Scott and Webley, Lisa C},
  shortjournal = {Information Visualization},
  volume = {20},
  pages = {192--204},
  doi = {10.1177/14738716211012608},
  year = {2021},
  journal = {Information Visualization}
}

@book{normanPsychologyEverydayThings1988,
  title = {The Psychology of Everyday Things},
  author = {Norman, Donald A.},
  publisher = {Basic books},
  isbn = {978-0-465-06709-1},
  year = {1988},
}

@incollection{graham2019intertextuality,
	author = {Allen, Graham},
	editor = {Lynch, Deidre Shauna},
	isbn = {9780197851470},
	title = {Intertextuality},
	year = {2019},
	booktitle = {Oxford Research Encyclopedia of Literature},
	publisher = {Oxford University Press},
	doi = {10.1093/acrefore/9780190201098.013.1072}
}

@article{brehmerMultiLevelTypologyAbstract2013,
  title = {A {{Multi-Level Typology}} of {{Abstract Visualization Tasks}}},
  author = {Brehmer, Matthew and Munzner, Tamara},
  doi = {10.1109/TVCG.2013.124},
  year={2013},
  volume={19},
  number={12},
  pages={2376-2385},
  journal = {IEEE Transactions on Visualization and Computer Graphics}
}

@inproceedings{munznerNestedModelVisualization2009,
  title = {A Nested Model for Visualization Design and Validation},
  booktitle = {{{IEEE Transactions}} on {{Visualization}} and {{Computer Graphics}}},
  author = {Munzner, Tamara},
  doi = {10.1109/TVCG.2009.111},
  volume={15},
  number={6},
  pages={921-928},
  year = {2009}
}

@article{lenz2025democratic,
	title={Democratic governance and policy complexity: revisiting the intelligence of democracy},
	author={Lenz, Alexa and Fern{\'a}ndez-i-Mar{\'\i}n, Xavier and Hinterleitner, Markus and Knill, Christoph and Steinebach, Yves},
	journal={Journal of European Public Policy},
	pages={1--28},
	year={2025},
	publisher={Taylor \& Francis}
}

@article{fowler2021implement,
	title={How to implement policy: Coping with ambiguity and uncertainty},
	author={Fowler, Luke},
	journal={Public Administration},
	volume={99},
	number={3},
	pages={581--597},
	year={2021},
	publisher={Wiley Online Library}
}

@article{dari2007uncertainty,
	title = {Uncertainty of Law and the Legal Process},
	author={Dari-Mattiacci, Giuseppe and Deffains, Bruno},
	pages={627--656},
	year={2007},
	volume = {163},
	ISSN = {0932-4569},
	DOI = {10.1628/093245607783242990},
	number = {4},
	journal = {Journal of Institutional and Theoretical Economics},
	publisher = {Mohr Siebeck},
}

@article{theil2025carefully,
	title={Carefully tailored: Doctrinal methods and empirical contributions},
	author={Theil, Stefan},
	journal={Oxford Journal of Legal Studies},
	doi = {10.1093/ojls/gqaf029},
	issn = {1464-3820},
	volume={45},
	number={4},
	pages={1047--1075},
	year={2025},
	publisher={Oxford University Press UK}
}

@book{baeza1999modern,
	title={Modern information retrieval},
	author={Baeza-Yates, Ricardo and Ribeiro-Neto, Berthier},
	volume={463},
	year={1999},
	publisher = {Addison Wesley},
	series    = {ACM Press}
}

@article{van2017concept,
  title = {On the Concept of Relevance in Legal Information Retrieval},
  author = {Van Opijnen, Marc and Santos, Cristiana},
  shortjournal = {Artif Intell Law},
  volume = {25},
  number = {1},
  pages = {65--87},
  doi = {10.1007/s10506-017-9195-8},
  year = {2017},
  journal = {Artificial Intelligence and Law},
  publisher = {Springer}
}

@article{staskoEvaluationSpacefillingInformation2000,
  title = {An Evaluation of Space-Filling Information Visualizations for Depicting Hierarchical Structures},
  author = {Stasko, John and Catrambone, Richard and Guzdial, Mark and Mcdonald, Kevin},
  doi = {10.1006/ijhc.2000.0420},
  year = {2000},
  journal = {International Journal of Human-Computer Studies},
  volume = {53},
  number = {5},
  pages = {663-694},
  issn = {1071-5819}
}

@article{collinsDocuBurstVisualizingDocument2009,
  title = {{{DocuBurst}}: {{Visualizing Document Content}} Using {{Language Structure}}},
  author = {Collins, Christopher and Carpendale, Sheelagh and Penn, Gerald},
  volume = {28},
  pages = {1039--1046},
  doi = {10.1111/j.1467-8659.2009.01439.x},
  year = {2009},
  journal = {Computer Graphics Forum}
}

@book{flyvbjerg2023big,
	title={How big things get done: The surprising factors that determine the fate of every project from home renovations to space exploration, and everything in between},
	author={Flyvbjerg, Bent and Gardner, Dan},
	publisher={Penguin Random House},
	year={2023}
}

@article{monteiro2009managing,
	title={Managing misunderstandings: The role of language in interdisciplinary scientific collaboration},
	author={Monteiro, Marko and Keating, Elizabeth},
	journal={Science communication},
	volume={31},
	number={1},
	pages={6--28},
	year={2009},
	publisher={SAGE Publications Sage CA: Los Angeles, CA}
}

@inproceedings{terzidou2023automated,
	title={Automated anonymization of court decisions: Facilitating the publication of court decisions through algorithmic systems},
	author={Terzidou, Kalliopi},
	booktitle={Proceedings of the Nineteenth International Conference on Artificial Intelligence and Law},
	pages={297--305},
	year={2023}
}

@article{coupette2018quantitative,
	title={Quantitative Rechtswissenschaft: Sammlung, Analyse und Kommunikation juristischer Daten},
	author={Coupette, Corinna and Fleckner, Andreas M},
	journal={JuristenZeitung},
	pages={379--389},
	year={2018},
	publisher={JSTOR}
}

@inproceedings{northAnalyticProvenanceProcess+interaction+insight2011,
  title = {Analytic Provenance: Process+interaction+insight},
  booktitle = {{{CHI}} '11 {{Extended Abstracts}} on {{Human Factors}} in {{Computing Systems}}},
  author = {North, Chris and Chang, Remco and Endert, Alex and Dou, Wenwen and May, Richard and Pike, Bill and Fink, Glenn},
  pages = {33--36},
  publisher = {ACM},
  doi = {10.1145/1979742.1979570},
  year = {2011},
}

@book{beyerContextualDesignDefining1998,
  title = {Contextual Design: Defining Customer-Centered Systems},
  editor = {Beyer, Hugh and Holtzblatt, Karen},
  date = {1998},
  publisher = {Morgan Kaufmann},
  isbn = {978-0-08-050304-2},
  year = {1998},
  address = {San Francisco, Calif}
}

@inproceedings{sperrleVIANAVisualInteractive2019,
  title = {{{VIANA}}: {{Visual Interactive Annotation}} of {{Argumentation}}},
  booktitle = {2019 {{IEEE Conference}} on {{Visual Analytics Science}} and {{Technology}} ({{VAST}})},
  author = {Sperrle, Fabian and Sevastjanova, Rita and Kehlbeck, Rebecca and El-Assady, Mennatallah},
  pages = {11--22},
  publisher = {IEEE},
  doi = {10.1109/VAST47406.2019.8986917},
  year = {2019},
}

@article{scheirerSenseConnectionAutomatic2016,
  title = {The Sense of a Connection: {{Automatic}} Tracing of Intertextuality by Meaning},
  author = {Scheirer, Walter and Forstall, Christopher and Coffee, Neil},
  shortjournal = {Digital Scholarship Humanities},
  volume = {31},
  pages = {204--217},
  doi = {10.1093/llc/fqu058},
  year = {2016},
  journal = {Digital Scholarship in the Humanities}
}

@article{tzanisGraphieNetworkbasedVisual2023,
  title = {Graphie: {{A}} Network-Based Visual Interface for the {{UK}}'s Primary Legislation},
  author = {Tzanis, Evan and Vivo, Pierpaolo and F\"orster, Yanik-Pascal and Gamberi, Luca and Annibale, Alessia},
  shortjournal = {F1000Res},
  volume = {12},
  pages = {236},
  doi = {10.12688/f1000research.129632.1},
  year = {2023},
  journal = {F1000Research}
}

@online{HistoryDoubleDiamond,
  title = {Eleven Lessons. {{A}} Study of the Design Process},
  author = {{Design Council}},
  url = {https://www.designcouncil.org.uk/fileadmin/uploads/dc/Documents/ElevenLessons_Design_Council%2520%25282%2529.pdf},
  urldate = {2026-09-11},
  year = {2003}
}

@article{sansoneLegalInformationRetrieval2022,
  title = {Legal {{Information Retrieval}} Systems: {{State-of-the-art}} and Open Issues},
  author = {Sansone, Carlo and Sperl\'i, Giancarlo},
  shortjournal = {Information Systems},
  volume = {106},
  pages = {101967},
  doi = {10.1016/j.is.2021.101967},
  year = {2022},
  journal = {Information Systems}
}

@inproceedings{lauLegalInformationRetrieval2005,
  title = {Legal Information Retrieval and Application to E-Rulemaking},
  booktitle = {Proceedings of the 10th International Conference on {{Artificial}} Intelligence and Law},
  author = {Lau, Gloria T. and Law, Kincho H. and Wiederhold, Gio},
  pages = {146--154},
  publisher = {ACM},
  doi = {10.1145/1165485.1165508},
  year = {2005}
}

@article{saravananImprovingLegalInformation2009,
  title = {Improving Legal Information Retrieval Using an Ontological Framework},
  author = {Saravanan, M. and Ravindran, B. and Raman, S.},
  shortjournal = {Artif Intell Law},
  volume = {17},
  pages = {101--124},
  doi = {10.1007/s10506-009-9075-y},
  year = {2009},
  journal = {Artificial Intelligence and Law}
}

@inproceedings{solovey2025interacting,
	author = {Solovey, Erin and Flanagan, Brian and Chen, Daniel},
	title = {Interacting with AI at Work: Perceptions and Opportunities from the UK Judiciary},
	year = {2025},
	isbn = {9798400713842},
	publisher = {Association for Computing Machinery},
	address = {New York, NY, USA},
	url = {https://doi.org/10.1145/3729176.3729192},
	doi = {10.1145/3729176.3729192},
	booktitle = {Proceedings of the 4th Annual Symposium on Human-Computer Interaction for Work},
	articleno = {5},
	numpages = {8},
	location = {
	},
	series = {CHIWORK '25}
}

@article{choi2024ai,
	title={AI assistance in legal analysis: An empirical study},
	author={Choi, Jonathan H and Schwarcz, Daniel},
	journal={Journal of Legal Education},
	volume={73},
	pages={384--420},
	year={2024},
	publisher={HeinOnline}
}

@article{martinho2025surveying,
	title={Surveying Judges about artificial intelligence: profession, judicial adjudication, and legal principles},
	author={Martinho, Andreia},
	journal={AI \& SOCIETY},
	volume={40},
	number={2},
	pages={569--584},
	year={2025},
	publisher={Springer}
}

@inproceedings{keimVisualAnalyticsDefinition2008,
  title = {Visual {{Analytics}}: {{Definition}}, {{Process}} and {{Challenges}}},
  booktitle = {Lecture {{Notes}} in {{Computer Science}}},
  author = {Keim, Daniel and Andrienko, Gennady and Fekete, Jean Daniel and G\"org, Carsten and Kohlhammer, J\"orn and Melan\c con, Guy},
  doi = {10.1007/978-3-540-70956-5_7},
  year = {2008}
}

@article{sachaKnowledgeGenerationModel2014,
  title = {Knowledge {{Generation Model}} for {{Visual Analytics}}},
  author = {Sacha, Dominik and Stoffel, Andreas and Stoffel, Florian and Kwon, Bum Chul and Ellis, Geoffrey and Keim, Daniel A.},
  shortjournal = {IEEE Trans. Visual. Comput. Graphics},
  volume = {20},
  pages = {1604--1613},
  doi = {10.1109/TVCG.2014.2346481},
  year = {2014},
  journal = {IEEE Transactions on Visualization and Computer Graphics}
}

@inproceedings{christensenDocumentscapeIntertextualitySequentiality2014,
  title = {Documentscape: Intertextuality, Sequentiality, \& Autonomy at Work},
  booktitle = {Proceedings of the {{SIGCHI Conference}} on {{Human Factors}} in {{Computing Systems}}},
  author = {Christensen, Lars Rune and Bjorn, Pernille},
  pages = {2451--2460},
  publisher = {ACM},
  doi = {10.1145/2556288.2557305},
  year = {2014}
}

@article{wangDefiningApplyingKnowledge2009,
  title = {Defining and Applying Knowledge Conversion Processes to a Visual Analytics System},
  author = {Wang, Xiaoyu and Jeong, Dong Hyun and Dou, Wenwen and Lee, Seok-Won and Ribarsky, William and Chang, Remco},
  volume = {33},
  pages = {616--623},
  doi = {10.1016/j.cag.2009.06.004},
  year = {2009},
  journal = {Computers \& Graphics}
}

@inproceedings{gomez2015understanding,
  title = {Understanding {{Large Legal Datasets}} through {{Visual Analytics}}},
  booktitle = {{{SIBGRAPI}} 2015-Conference on Graphics, Patterns and Images},
  author = {Gomez-Nieto, Erick and Casaca, Wallace and Hartmann, Ivar and Nonato, Luis Gustavo},
  volume = {4},
  year = {2015}
}

@article{bokwonleeNetworkStructureReveals2018,
  title = {Network {{Structure Reveals Patterns}} of {{Legal Complexity}} in {{Human Society}}: {{The Case}} of the {{Constitutional Legal Network}}},
  author = {{Bokwon Lee} and Lee, Bokwon and {Kyumin Lee} and Lee, Kyu-Min and {Jae-Suk Yang} and Yang, Jae-Suk},
  doi = {10.2139/ssrn.3202925},
  year = {2018}
}

@article{mentzingenUnveilingLegalComplexity2025,
  title = {Unveiling Legal Complexity: A Systematic Review on the Visual Analytics of Legal Corpora},
  author = {Mentzingen, Hugo and Ant\'onio, Nuno and Bacao, Fernando},
  pages = {1--36},
  doi = {10.1080/13600869.2025.2497630},
  year = {2025},
  journal = {International Review of Law, Computers \& Technology}
}

@inproceedings{lettieriAffordanceLawSliding2020,
  title = {The {{Affordance}} of {{Law}}. {{Sliding Treemaps}} Browsing {{Hierarchically Structured Data}} on {{Touch Devices}}},
  booktitle = {2020 24th {{International Conference Information Visualisation}} ({{IV}})},
  author = {Lettieri, Nicola and Guarino, Alfonso and Malandrino, Delfina and Zaccagnino, Rocco},
  pages = {16--21},
  publisher = {IEEE},
  doi = {10.1109/IV51561.2020.00013},
  year = {2020},
}

@inproceedings{endertSemanticInteractionVisual2012,
  title = {Semantic {{Interaction}} for {{Visual Text Analytics}}},
  booktitle = {Proceedings of the {{SIGCHI Conference}} on {{Human Factors}} in {{Computing Systems}}},
  author = {Endert, Alex and Fiaux, Patrick and North, Chris},
  pages = {473--482},
  publisher = {ACM},
  doi = {10.1145/2207676.2207741},
  year = {2012},
}
	\balance
	
	\clearpage
\onecolumn
\appendix
\crefalias{section}{appendix}

\section*{Supplementary Material for \papertitle}\label{appendix:appendix}

In this appendix, we collect supplementary materials elaborating on the content presented in our main paper. 
Specifically, we offer the following materials. 

\begin{enumerate}[label={}]
	\item \Cref{appendix:legal-work}:~\nameref{appendix:legal-work}
	\item \Cref{appendix:extended-related-work}:~\nameref{appendix:extended-related-work}
	\item \Cref{appendix:participant-demographics-table}:~\nameref{appendix:participant-demographics-table}
	\item \Cref{appendix:requirements}:~\nameref{appendix:requirements}
	\item \Cref{appendix:design-process}:~\nameref{appendix:design-process}
	\item \Cref{appendix:interface-details}:~\nameref{appendix:interface-details}
	\item \Cref{appendix:prototype-evaluation}:~\nameref{appendix:prototype-evaluation}
\end{enumerate}

\clearpage

	\section{A Primer on Legal Systems}\label{appendix:legal-work}

All legal work happens inside \emph{legal systems}.
As the plural suggests, there exist many legal systems, 
and they can be nested inside each other to form \emph{multi-level legal systems}, such as in the case of federalist nations (e.g., the United States or Germany) or the European Union (created by its member states). 
In legal systems,
individuals or institutions create and interpret legal documents,
continuously producing new outputs that feed back into the system and combine into a growing corpus of rules,
i.e., decisions at different levels of abstraction, persisted as texts \cite{coupette2021measuring,punder2026power}. 

The legal work yielding these texts is inherently \emph{heterogeneous}. 
\emph{Legislators} draft and enact new legislation,
directly changing the state of the law. 
\emph{Administrative officials} implement legislation and enforce it in individual cases. 
\emph{Judges} decide on cases brought
to the courts, 
interpreting the \emph{abstract} law
and applying it to \emph{concrete} real-life scenarios.
While a decision directly affects only the parties involved,
as \emph{precedent},
it can also impact future cases (the details depend on the jurisdiction in question).
\emph{Attorneys} guide their clients through the legal system,
advising them on their legal options, 
shaping their future by designing contracts, 
or litigating their past before the courts.
\emph{Legal scholars} guide this process from a meta perspective, 
analyzing and organizing legal content, 
detecting uncertainties and discussing ambiguities, 
and deriving what \emph{is} and what \emph{should be} the law.
Depending on the jurisdiction,
their opinions may influence the outputs produced by other actors as well. 

Despite their distinct roles and objectives,
all actors in legal systems share a primary mode of working:
\emph{interaction with text}.
\emph{Legal text} constitutes the base material of artifacts in legal systems and the core means of legal action, 
and it features three peculiar properties: 
\emph{heterogeneous dynamics}, \emph{granular connectivity}, and \emph{normative intertextuality} (see \Cref{sec:requirements:abstraction}).
As illustrated in \Cref{fig:legal-work}, these properties produce the specific interaction requirements of legal work that collectively distinguish it from work in other text-based domains.

	\section{Extended Related Work}\label{appendix:extended-related-work}

The following section supplements the discussion of related work and background in \Cref{sec:related_work} by providing a broader overview of the relevant literature for additional context.

\subsection{Models of Sense-Making and Visual Analytics}

Models of sense-making~\cite{cook2005illuminating, pirolli2005sensemaking, kleinMakingSenseSensemaking2006a} are closely related to models for information visualization~\cite{shrinivasanSupportingAnalyticalReasoning2008} and Visual Analytics~(VA)~\cite{keimVisualAnalyticsDefinition2008, sachaKnowledgeGenerationModel2014}.
On the abstract end of the spectrum, the \emph{Data/Frame theory} can be applied to everyday situations~\cite{kleinMakingSenseSensemaking2006a}, where models for VA specifically posit the advantages of data visualization for sense-making and incorporate aspects of perception not considered elsewhere.
However, their level of abstraction is still too high for providing guidance on interface design, as they, too, are purposed to analyze the sense-making process by not committing to a specific domain or abstraction thereof.
Later models started to touch on the importance of treating knowledge more differentially, distinguishing between implicit and explicit, or render implicit knowledge explicitly by the use of a VA system~\cite{wangDefiningApplyingKnowledge2009}, which is closely related to the idea of enriching the analytical process with provenance~\cite{northAnalyticProvenanceProcess+interaction+insight2011, kaixuAnalyticProvenanceSensemaking2015}.
The conversion of knowledge is also a need that came up with our participants in the prototype evaluation~(see~\Cref{sec:evaluation:results:g3}).

\subsection{Adaptive Meaning Construction in Text-Heavy Domains}

User intents of Adaptive Meaning Construction~(AMC) also play a role in political discourse analysis~\cite{sperrleVIANAVisualInteractive2019}, 
history~\cite{scheirerSenseConnectionAutomatic2016}, scientific literature exploration, or software engineering~\cite{horvathUsingAnnotationsSensemaking2022}.
In scientific literature exploration, prior work succeeds in reducing the distance between data points within a source~\cite{fokQlarifyRecursivelyExpandable2024}, contextualizing different data sources~\cite{kimDataDiveSupportingReaders2024, duckFindingNeedlesDocument2025}, and supporting the synthesis of different sources~\cite{kangSynergiMixedInitiativeSystem2023, choeCrossLitConnectingVisual2026}.
For improved understanding of scientific abstracts, \citeauthor{fokQlarifyRecursivelyExpandable2024} co-locate information from the full paper on demand directly inside of the abstract, reducing the distance between data points~\cite{fokQlarifyRecursivelyExpandable2024}.
With \emph{NEEDLE}, \citeauthor{duckFindingNeedlesDocument2025} provide an interface for effective scientific claim retrieval and contextualization through search and close-reading supported by a canvas for consolidating retrieved claims~\cite{duckFindingNeedlesDocument2025}.
\citeauthor{choeCrossLitConnectingVisual2026} leverage the modalities of text and visualization to support the visual synthesis of different data sources into the text of a scientific manuscript~\cite{choeCrossLitConnectingVisual2026}.
\citeauthor{endertSemanticInteractionVisual2012} have abstracted from specific types of text to propose a design space for semantic interaction in visual text analytics~\cite{endertSemanticInteractionVisual2012}.
For the characterization of text-heavy domains, prior research on collaborative work environments has primed the interplay between document collections, dynamics, connectivity, and intertextuality but did not investigate a domain-agnostic abstraction~\cite{christensenDocumentscapeIntertextualitySequentiality2014}.

\subsection{Visual Analytics for Law}

The AMC-characterizing properties of \emph{heterogeneous dynamics}, \emph{granular connectivity}, and \emph{normative intertextuality}~(see~\Cref{sec:requirements:abstraction}) have yet to take center stage for the intersection of VA and law.
While the necessity for interactivity, hierarchy, and interpretability are being recognized, a cohesive framework or taxonomy is missing~\cite{mclachlanVisualisationLawLegal2021, mentzingenUnveilingLegalComplexity2025}.
As such, past works in visualization for law did not account for AMC and consequently fell short in respecting required granularity, for example, providing only document-level node-link diagrams, which are not sufficient to understand influences on legal meaning.
Instead, existing work focuses on \emph{how} to visualize legal information using different techniques, such as node-link diagrams, timelines, concept maps, hierarchical representations, and even geospatial visualizations~\cite{mentzingenUnveilingLegalComplexity2025}.
For example, existing work focuses on rendering legal references as node-link diagrams~\cite{lettieriLegalMacroscopeExperimenting2017, bokwonleeNetworkStructureReveals2018, lacavaLawNetVizWebbasedSystem2022}, interactively exploring document relationships~\cite{resckLegalVisExploringInferring2023}, and expressing legal hierarchies~\cite{gomez2015understanding, lettieriAffordanceLawSliding2020}.
However, these techniques are usually isolated and do not come as a holistic approach.

\clearpage

	\clearpage
\onecolumn

\section{Participant Demographics Details}\label{appendix:participant-demographics-table}

We conducted two user-focused evaluations:
a requirements elicitation (RE, see \Cref{sec:requirements}) and a prototype evaluation (PE, see \Cref{sec:evaluation-design} and \Cref{sec:prototype-evaluation}).
The detailed overview of our participants is shown in  \Cref{tab:participant-demographics}.

\begin{table*}[h]
	\caption{%
		Demographic details for the 30 participants in our studies for Requirements Elicitation (RE) and Prototype Evaluation (PE). 
		Five experts participated in both requirements elicitation and prototype evaluation, such that the survey that was part of the prototype-evaluation study also served to confirm the needs distilled from our requirements-elicitation study.
	}
	\Description{The table shows demographic details for the 30 participants in our studies for Requirements Elicitation (RE) and Prototype Evaluation (PE). Five experts participated in both requirements elicitation and prototype evaluation, such that the survey that was part of the prototype-evaluation study also served to confirm the needs distilled from our requirements-elicitation study.} %
	\label{tab:participant-demographics}
	\begin{tblr}{
		colspec={
			r
			c
			l
			l
			Q[l,wd=6cm]
			r
			r
			r
			l %
			c
			c
			c
		},
		rowsep=.75pt,
		column{1} = {leftsep=0pt, rightsep=0.75em},
		column{2} = {leftsep=0pt, rightsep=0.75em},
		column{3} = {leftsep=0pt, rightsep=0.75em},
		column{4} = {leftsep=0pt, rightsep=0.5em},
		column{5} = {leftsep=0pt, rightsep=0.5em},
		column{6} = {leftsep=0pt, rightsep=0.5em},
		column{7} = {leftsep=0pt, rightsep=0.5em},
		column{8} = {leftsep=0pt, rightsep=0.5em},
		column{9} = {leftsep=-0.25em, rightsep=-0.25em},
		column{10} = {leftsep=0pt, rightsep=0.5em},
		column{11} = {leftsep=0pt, rightsep=0pt},
	}
	\toprule
	
	&&&&&
	\SetCell[c=3]{c} \textbf{Years Exp.$^\ast$}
	&&
	&
	&
	\SetCell[c=2]{c} \textbf{Study}
	&
	\\
	
	\cmidrule{6-8}
	\cmidrule{10-11}
	
	\bfseries PID
	& \bfseries Age
	& \bfseries \Hermaphrodite
	& \bfseries Main Actor Type
	& \bfseries Current Role
	& \bfseries Law
	& \bfseries EU
	& \bfseries Viz
	& \phantom{abc}
	& \bfseries RE
	& \bfseries PE
	\\
	
	\midrule
	
	1&[25,30)&\Male&scholarly&PhD Researcher&7& 2&--&&\checkmark&\\ %
	2&[35,40)&\Male&scholarly&Professor, Legal Tech Entrepreneur&17&16&20&&\checkmark&$\circ^\ddagger$\\ %
	3&[35,40)&\Female&scholarly&Professor, Part-Time Judge&17 &11&0&&\checkmark&$\circ^\dagger$\\ %
	4&[50,55)&\Female&scholarly&Senior Researcher&23 &23&--&&\checkmark&\\ %
	5&[45,50)&\Male&institutional&Public Servant&25 &15&10&&\checkmark&\checkmark\\ %
	
	\midrule[dotted]
	
	6&[35,40)&\Male&societal&Legal Tech Entrepreneur, former Lawyer&14& 14&--&&\checkmark&\\ %
	7&[25,30)&\Male&scholarly&PhD Researcher&8 &5&--&&\checkmark&\\ %
	8&[45,50)&\Male&scholarly&Professor&25 &25&--&&\checkmark&\\ %
	9&[40,45)&\Male&institutional&Public Servant and Senior Researcher&20&17&10&&\checkmark&\checkmark\\ %
	10&[25,30)&\Male&scholarly&PhD Researcher, former Lawyer&8 &5&--&&\checkmark&\\ %
	
	\midrule[dotted]
	
	11&[30,35)&\Male&scholarly&PhD Researcher&10&8& 10&&\checkmark&\checkmark\\ %
	12&[25,30)&\Male&scholarly&PhD Researcher&4&3&4&&\checkmark&\checkmark\\ %
	13&[30,35)&\Female&scholarly&PhD Researcher, former Lawyer&5&1&2&&\checkmark&\checkmark\\ %
	14&[30,35)&\Female&institutional&Public Servant&5&5&--&&\checkmark&\\ %
	15&[30,35)&\Female&societal&Head of Policy in a Civil-Society Organization&11&4&--&&\checkmark&\\ %
	
	\midrule[dotted]
	
	16&[30,35)&\Male&scholarly&PhD Researcher, former Lawyer&10 &5&0& &&\checkmark\\ %
	17&[35,40)&\Female&scholarly&PhD Researcher, former Lawyer&8 &3&0& &&\checkmark\\ %
	18&[40,45)&\Male&scholarly&Professor&20&15&0& &&\checkmark\\ %
	19&[25,30)&\Male&scholarly&PhD Researcher&8 &3&0& &&\checkmark\\ %
	20&[25,30)&\Female&scholarly&PhD Researcher&8& 2&0& &&\checkmark\\ %
	
	\midrule[dotted]
	
	21&[25,30)&\Female&scholarly&PhD Researcher&9& 7&0& &&\checkmark\\ %
	22&[30,35)&\Male&scholarly&PhD Researcher&4&6&0&&&\checkmark\\ %
	23&[30,35)&\Female&scholarly&PhD Researcher&2&2&0&&&\checkmark\\ %
	24&[80,85)&\Male&societal&Civil-Society Activist, former Public Servant&57&37&1& &&\checkmark\\ %
	25&[55,60)&\Male&scholarly&Professor&34&30&25&&&\checkmark\\ %
	
	\midrule[dotted]
	
	26&[40,45)&\Female&scholarly&Professor&16&20&3&&&\checkmark$^{\ast\ast\textbackslash\ast\ast\ast}$\\ %
	27&[40,45)&\Male&scholarly&Editor and Researcher&25&25&0&&&\checkmark\\ %
	28&[25,30)&\Female&scholarly&PhD Researcher&7&3&2& &&\checkmark\\ %
	29&[25,30)&\Male&scholarly&PhD Researcher&6&3&3& &&\checkmark\\ %
	30&[45,50)&\Male&institutional&Public Servant&20&20&5& &&\checkmark$^{\ast\ast}$\\ %

	\bottomrule
\end{tblr}

	\vspace*{0.5em}
	\raggedright
	\emph{$\ast:$ Participants made different assumptions when answering the questions about their years of experience.}
	
	\emph{$\ast\ast:$ Evaluation interview was a shorter in duration.}
	
	\emph{$\ast\ast\ast:$ Due to shorter duration of evaluation, data on `Efficiency \& Usability' is missing.}
	
	\emph{$\dag:$ Participated in an earlier draft of our PE study that led us to reconsider our evaluation approach (see \cref{sec:evaluation-design}).}
	
	\emph{$\ddag:$ Tested our revised PE setup following the experience reported in $\dag$, confirming ecological validity (see \cref{sec:evaluation-design}).}
\end{table*}

\clearpage

	\section{Requirements Elicitation Details}\label[appendix]{appendix:requirements}

In the following, we supplement the discussion of our requirements elicitation in \Cref{sec:requirements}, providing the demographic questions asked (\Cref{appendix:requirements-demographic-questions}) and the interview guide we followed (\Cref{appendix:requirements-interview-guide}).
For the detailed participant information, see the combined \Cref{tab:participant-demographics} in \Cref{appendix:participant-demographics-table}, referring to the Requirements Elicitation (RE) column.

\subsection{Demographic Questionnaire}\label[appendix]{appendix:requirements-demographic-questions}
We collected the demographic background of participants separately before the interview recording.
The identifying information, where needed, is stored apart from the analysis dataset.

\begin{itemize}
	\item What is your participant ID? \textit{(assigned; name held separately on the consent form)}
	\item What is your age? \textit{(exact / 5-year-bands / prefer not to disclose)}
	\item Which gender do you identify with? \textit{(female / male / non-binary / prefer to self-describe / prefer not to disclose)}
	\item What group do you consider yourself to belong to? \textit{(scholarly / societal / institutional)}
	\item What are your previous and current roles? \textit{(free text)}
	\item How many years of experience do you have \dots
		\begin{itemize}
			\item \dots in general? \textit{(exact)}
			\item \dots with European legal documents? \textit{(exact)}
		\end{itemize}
\end{itemize}

\subsection{Interview Guide}\label[appendix]{appendix:requirements-interview-guide}
\paragraph{1. Opening \& Context}
\begin{enumerate}[label=(\alph*)]
	\item What is your background?
	\begin{itemize}
		\item \textit{Probe:} Did you go through legal training (if so, in which country or countries)? Do you have training in other disciplines? How did you come to this work?
	\end{itemize}
	\item What is the context of your work, and why do you need to interact with European legal documents?
	\begin{itemize}
		\item \textit{Probe (scholarly):} Do you primarily do research or teaching? What are your focus areas (e.g., specific subdomains of EU law, comparative law, law and economics, ...) and methods (e.g., doctrinal analysis, qualitative or quantitative empirical methods, socio-legal methods, computational methods)?
		\item \textit{Probe (societal):} What is your primary activity? For example, advocacy or monitoring?
		\item \textit{Probe (institutional):} What is your primary activity? For example, compliance or drafting amendments?
	\end{itemize}
	\item What do you typically produce as an output of your work, and who is the typical audience of that work?
	\begin{itemize}
		\item \textit{Probe (scholarly):} For example, journal articles, books, conference contributions, or policy recommendations? Addressed to whom?
		\item \textit{Probe (societal):} For example, a memo, a position paper, or advocacy material? Addressed to whom?
		\item \textit{Probe (institutional):} For example, an amendment proposal or a compliance checklist? Addressed to whom?
	\end{itemize}
	\item At what level(s) do you work: European Union, international, national, or subnational?
	\begin{itemize}
		\item \textit{Probe:} How does the interaction of legal documents at different levels affect your work? 
	\end{itemize}
	\item How do you typically access legal information?
	\begin{itemize}
		\item \textit{Probe:} Which information systems, databases, or other tools do you normally use?
	\end{itemize}
\end{enumerate}

\paragraph{2. Concrete Walkthrough}\hspace{1em}
	Talk to me about a recent time you analyzed a specific EU legal document.
	Why were you conducting the analysis?
	What were you trying to achieve and what did you do, step by step?
	\begin{itemize}
		\item \textit{Probe:} If you have it open, can you show me?
		\item \textit{Probe:} Where did the document come from? Did you know whether it was the current version? How?
		\item \textit{Probe:} What was the hardest or most time-consuming part?
		\item \textit{Probe:} Did anyone else get involved? How did you share or hand off the result?
		\begin{itemize}
			\item \textit{Probe:} Do you perform this kind of activity mostly alone or as part of a collaborative setting?
		\end{itemize}
		\item \textit{Probe:} How often do you perform this kind of activity?
		\item \textit{Probe:} Roughly, how many documents does such activity involve?
	\end{itemize}

\paragraph{3. General Workflow}
\begin{enumerate}[label=(\alph*)]
	\item Thinking more generally, what workflows do you rely on in your work?
	\begin{itemize}
		\item \textit{Probe:} What is the motivation behind these workflows?
		\item \textit{Probe:} Do you perform this kind of activity mostly alone or as part of a collaborative setting?
	\end{itemize}
	\item How do you know which (legal) documents to look at?
	\begin{itemize}
		\item \textit{Probe:} What types of (legal) documents do you consider for your work?
	\end{itemize}
	\item How do you navigate within and across (legal) documents?
	\begin{itemize}
		\item \textit{Probe:} How do you trace cross-references and identify related documents?
		\item \textit{Probe:} How do you connect different documents?
	\end{itemize}
	\item When you interact with a document, what are you looking for? And what underlying questions are you ultimately trying to answer?
	\begin{itemize}
		\item \textit{Probe:} What are the relevant legal entities that you are looking for? For example, norms, definitions, doctrinal arguments, recitals, or judicial interpretations.
	\end{itemize}
	\item  Do you track which version of a text is current (or currently relevant to you)? If so, how?
	\begin{itemize}
		\item \textit{Probe:} Specifically, how do you handle proposals, amendments, and consolidated texts? What about versions from the trilogue stages?
		\item \textit{Probe:} Do you also consider older versions? If so, how and why?
		\item \textit{Probe:} Have you ever accidentally worked with the wrong version?
	\end{itemize}
	\item Do you compare documents directly? If so, how and why?
	\begin{itemize}
		\item \textit{Probe:} How do you find out what changed? What types of changes are you interested in?
		\item \textit{Probe:} What is the unit of change that you consider? For example, articles, paragraphs, sentences, or words.
		\item \textit{Probe:} How do you know which of the changes are legally significant?
	\end{itemize}
	\item Do you work across languages?
	\begin{itemize}
		\item \textit{Probe:} How do you handle documents in multiple official languages? Do you also compare them?
	\end{itemize}
	\item As you work through a document, how and where do you record your findings?
	\begin{itemize}
		\item \textit{Probe:} Do you keep what you've worked on for later reuse? How do you find it again?
	\end{itemize}
	\item What tools do you use to support these workflows, and why those?
	\begin{itemize}
		\item \textit{Probe:} Do you use analogous tools such as pen \& paper, (page) markers, post-its or whiteboards?
		\item \textit{Probe:} Do you use digital tools like information systems or databases?
		\item \textit{Probe:} Does AI play a role in your current work? Why do you (or do you not) use it, and to which ends?
		\begin{itemize}
			\item \textit{Probe:} Is the use of AI restricted to certain parts of your workflows? If so, how and why?
		\end{itemize}
	\end{itemize}
\end{enumerate}

\paragraph{4. Pain Points \& What Works}
\begin{enumerate}[label=(\alph*)]
	\item What works well in your current setup that you would not want to lose?
	\item What challenges or frustrations do you face with your tools and workflows?
	\begin{itemize}
		\item \textit{Probe:} Are there any examples that you remember when tools or workflows failed you?
		\item \textit{Probe:} Any workarounds that you have developed?
		\item \textit{Probe:} How do you know that you can trust a source or tool?
		\item \textit{Probe:} How--if at all--do you verify that the information you get from a tool is correct?
	\end{itemize}
\end{enumerate}

\paragraph{5. Envisioned Support}
\begin{enumerate}[label=(\alph*)]
	\item If you had a magic wand, what kind of support would you wish for and why?
	\item Are there tools, techniques, or approaches from other parts of your work or life that you wish you could apply to your workflows?
	\begin{itemize}
		\item \textit{Probe:} How might that integrate with the tools you already use?
	\end{itemize}
	\item What would make you adopt, not adopt, or abandon a tool in your work?
	\item Beyond reading documents one by one, are there other ways in which you would like to see or interact with legal documents that would help you in your work?
	\begin{itemize}
		\item \textit{Probe:} Are there questions in your work that could benefit from a quantitative perspective – how often something occurs, how much, how many – rather than the meaning of a single text? If so, why, and how do you handle those questions at the moment?
		\item \textit{Probe:} Do you ever feel the need to step back from individual documents to see broader patterns across many of them? If so, why, and how do you approach this task at the moment?
		\item \textit{Probe:} When you are making sense of how documents relate, do you ever sketch it out, draw diagrams, or wish you could see the relationships laid out visually? If so, why? Do you have any specific visualizations in mind that would help you?
	\end{itemize}
\end{enumerate}

\paragraph{6. Conclusion}
\begin{enumerate}[label=(\alph*)]
	\item Of everything we discussed, what costs you the most time or worries you the most?
	\item Is there anything else related to this interview that you would like to share with us?
	\item Are you willing to follow up with us on the requirements that we distilled from our interviews in another meeting?
	\item Is there someone else you would recommend that we speak to, or who might be interested in evaluating the tool we will build based on our interviews?
\end{enumerate}

\subsection{Elicited Requirements}\label[appendix]{appendix:extended-requirements}

In the following paragraphs, we categorize and present the elicited requirements in more detail, referencing the sources of our insights via the participant IDs specified in \Cref{tab:participant-demographics}.

\par{\textbf{Data Access.}}
Discovering and collecting relevant material
is a core prerequisite for deriving meaning.
In this regard,
lawyers need \emph{seamless} and \emph{contextualized} access to legal texts. %
Seamless points toward the reduction of friction,
e.g., a \emph{quick} and \emph{easy} access via various identifiers (\requirementsExpertOne, \requirementsExpertEight),
or \emph{coherently structured} data responses in terms of properties and metadata (\requirementsExpertTwo).
Contextualized means \emph{encompassing varying types of documents} such as legislation and jurisprudence ~(\requirementsExpertSeven)
as well as \emph{enriching} a data response with \emph{metadata},
such as \emph{temporal versions} of the queried document or other \emph{related documents} (\requirementsExpertOne, \requirementsExpertTwo, \requirementsExpertFour, \requirementsExpertFive, \requirementsExpertSix).

\par{\textbf{Search and Navigation.}} 
\emph{Seamless} and \emph{contextualized} data access
should be \emph{operationalized} and \emph{implement} in search and navigation functionalities.
This covers the query itself as well as `loaded' responses that facilitate the next step in legal work.
Lawyers need reliable semantic search~(\requirementsExpertTwo, \requirementsExpertSeven, \requirementsExpertEight, \requirementsExpertNine, \requirementsExpert{12}, \requirementsExpert{13}).
Participants mentioned a need for expressive and fine-grained control over the query,
e.g., \emph{filtering} the search pool by leveraging document metadata (\requirementsExpertTwo, \requirementsExpertSeven, \requirementsExpertEight, \requirementsExpert{13}),
\emph{extending} it by including documents across sources and languages~(\requirementsExpert{13}),
or \emph{combining} several search terms to detect correlating documents~(\requirementsExpertOne, \requirementsExpertEight).
At the same time,
lawyers desire navigation support to reduce manual query operations.
At a high level, this means that an interface should allow easy access to linked documents~(\requirementsExpertFour, \requirementsExpertSeven, \requirementsExpert{14}, \requirementsExpert{15}).
A lower-level facet of this may be found in exposing relationships as clickable hyperlinks in the actual text,
rather than in a detached list \emph{out of the current context}~(\requirementsExpertFour, \requirementsExpert{14}).

\par{\textbf{Versioning and Comparison.}}
Legal documents are subject to versioning (although mostly provided without proper version control).
This primarily affects the \emph{temporal} dimension,
i.e., how legal acts change over time.
Participants wished for a place to \emph{track} such changes~(\requirementsExpertOne, \requirementsExpertSeven)
and \emph{compare} different document versions~(\requirementsExpertTwo, \requirementsExpertFour).
This is particularly relevant in the context of \emph{consolidated versions}~(\requirementsExpertFour, \requirementsExpertSeven, \requirementsExpertTen),
which supplement legislative acts that are often only published \emph{once} in their full version and then amended by \emph{other} documents.
The consolidated version is a derived state of the initial (outdated) version with all change instructions applied.
In other words, the legal document landscape is a changing,
dynamic environment that calls for comparison modes in legal interfaces.
In order to efficiently compare different documents,
many participants would feel supported by presenting document next to each other
and viewing them \emph{in parallel},
rather than having to \emph{sequentially} switch back and forth between them~(\requirementsExpertFour, \requirementsExpertEight, \requirementsExpertTen, \requirementsExpert{11}, \requirementsExpert{12}).

\par{\textbf{Relationships and Context.}}
To adequately incorporate intertextuality,
lawyers cannot afford to study a text in isolation.
Instead,
they need to consider related documents that may \emph{specify} or even \emph{change} the meaning of an initial document.
As a first step, this means making context \emph{visible},
e.g.,
\emph{presenting} all related documents in a separate tab~(\requirementsExpertOne)
or indicating special kinds of relations, such as national law transposing EU legislation~(\requirementsExpertTen).
The next step beyond visibility is to ensure that context is readily \emph{accessible}.
In this spirit, several participants wished for a way to \emph{explore} interconnected documents~(\requirementsExpertTwo, \requirementsExpertTen, \requirementsExpert{12}, \requirementsExpert{13}).

\par{\textbf{Summarization and Visualization.}} %
Going beyond the support in \emph{navigation},
an interface may also support lawyers in the process of meaning \emph{adaptation}.
We categorize needs based on how they build on the underlying data.
Perceiving law \emph{as text},
participants \requirementsExpertOne, \requirementsExpertNine, and \requirementsExpertTen suggested \emph{summaries}
to both \emph{accelerate comprehension}
and enable quick \emph{disposal of irrelevant documents}.
Adopting a \emph{quantiative} perspective,
\requirementsExpert{12} asked for insights about the reference frequency of judgments or legislation.
Furthermore, many participants reported their mental model to be organized like a map that connects multiple related document~(\requirementsExpertTen, \requirementsExpert{12}, \requirementsExpert{13}).
These maps are shaped by hierarchical relations between concepts and documents~(\requirementsExpertThree, \requirementsExpert{12}, \requirementsExpert{13}).
Ultimately,
this corresponds to an interpretation of legal systems \emph{as networks}.
To support lawyers in the comprehension phase, 
quantitative and structural data present visualization opportunities that advance interactions beyond the mere presentation of text.
In this regard, participants explicitly mentioned networks, timelines, and hierarchical visualizations such as trees~(\requirementsExpertOne, \requirementsExpert{12}, \requirementsExpert{13}).

\par{\textbf{Usability and Performance.}}
Lastly,
participants expressed desiderata regarding the \emph{usability} and \emph{performance} of any legal interface.
In the context of EU law, some experts expected the typesetting not to deviate from its presentation in the \emph{Official Journal}~(\requirementsExpertTen, \requirementsExpert{11}).
\requirementsExpertFive~formulated the higher-level objective of fixing the underlying problem of \emph{information overload}.
This was echoed by \requirementsExpertTen~and \requirementsExpert{13}, who wished for \emph{clarity}, especially with regard to search results.
Further along these lines,
\requirementsExpertThree~mentioned that they would benchmark a tool based on its ability to actually reduce their work.

\clearpage

	\section{Design Process Details}\label{appendix:design-process}
\FloatBarrier

In this section, we gather additional materials from our design process.
This includes early sketch drawings of the three different views and their modes (\Cref{fig:appendix-skwtch-three-views}), an initial annotated sketch of the single-document view interface (\Cref{fig:appendix-sketch-single-document-view}), 
and a screenshot of the custom annotation tool we used for coding our interviews (\Cref{fig:appendix-annotation-tool}).

\begin{figure}
	\includegraphics[width=0.75\textwidth]{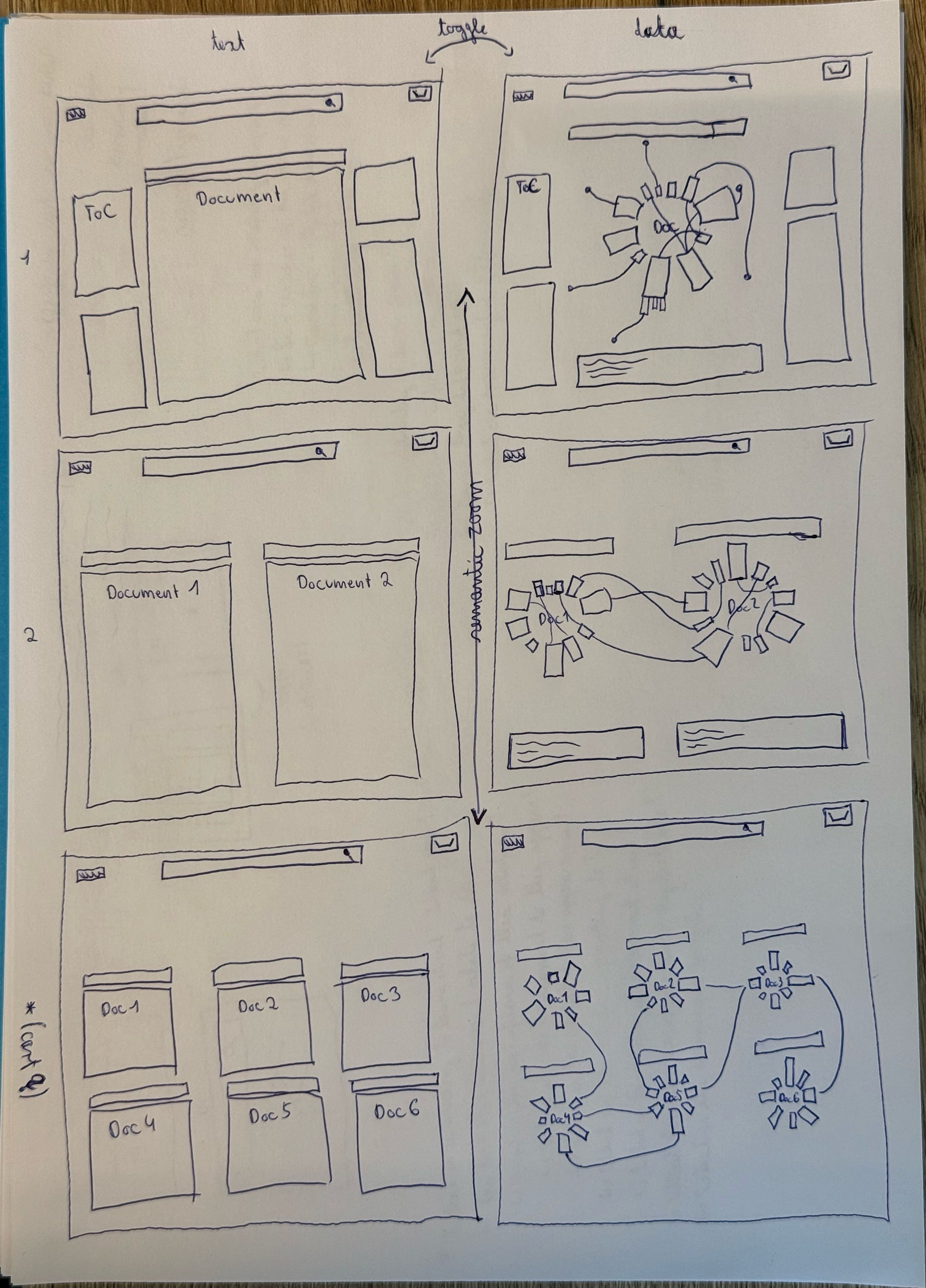}
	\caption{Sketch of the three views (single-document, two-documents, and many-documents) in text and data mode. The all-documents view is not depicted here.}
	\label{fig:appendix-skwtch-three-views}
	\Description{Shown is an abstract representation of the two different exploration modes (text-focused and data-focused) and their corresponding views for single-document, two-document, and many-documents.
		The single document view in text mode shows the main text centered and highlighted for context, the table of contents and incoming and outgoing references on the left top and bottom receptively, and context information on the right.
		In data-mode, is shows a radial glyph highlighting the relations and references of the document.
		The two-documents view in text mode shows two text side-by-side and also allows for git-like difference computation to highlight changes, e.g., between versions.
		Several different display options are available, like side-by-side, inline, or condensed.
		In data-mode, it shows two glyphs side-by-side, while connections (referenced) between them are displayed as bundled links.
		The many-documents view extends these functionalities to multiple documents. } %
\end{figure}

\begin{figure}
	\includegraphics[width=.8\textwidth]{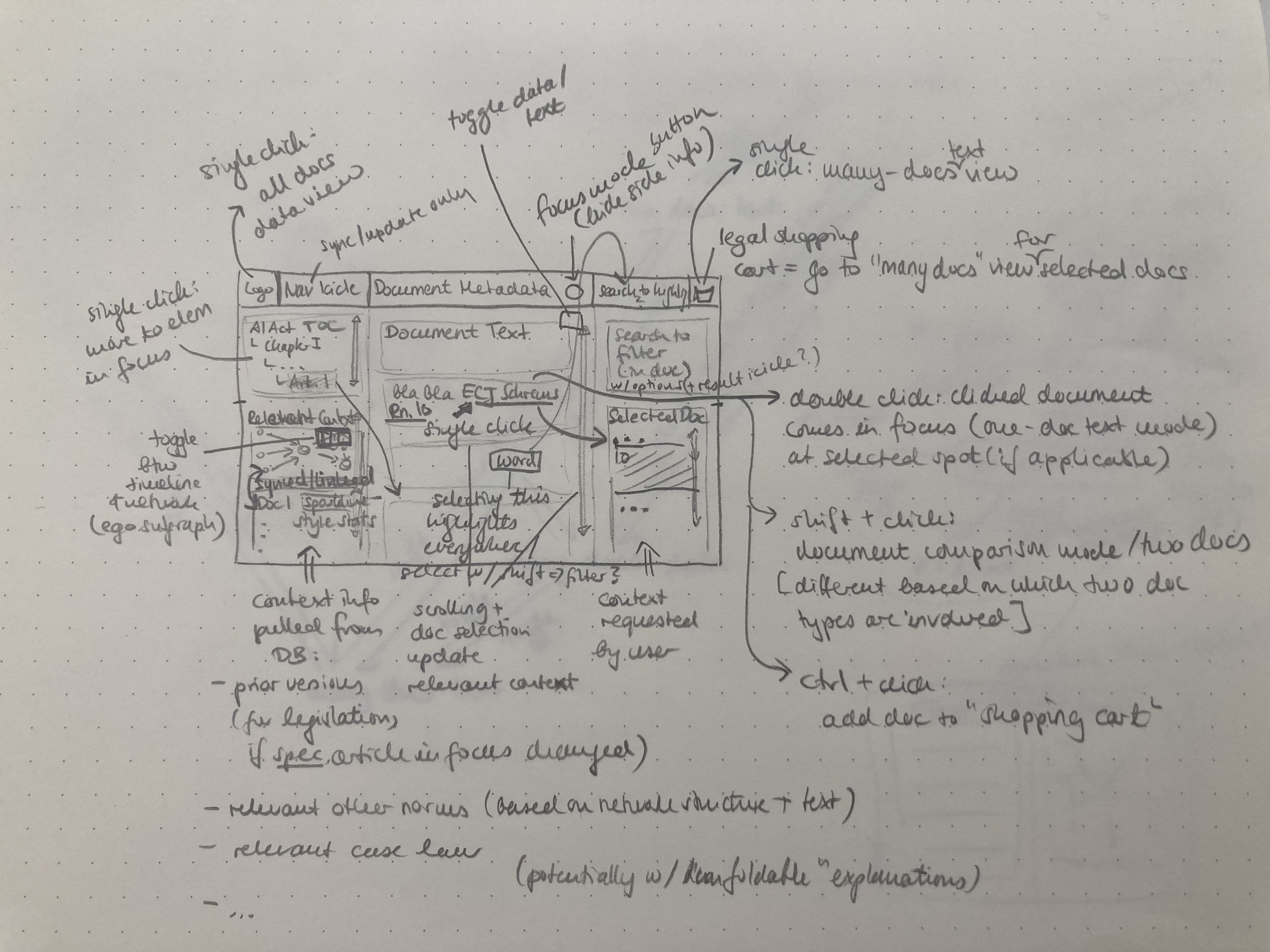}
	\caption{Sketch of the single-document view in text mode.}
	\label{fig:appendix-sketch-single-document-view}
	\Description{The sketch shows the single-document view in text mode.
		The main text is centered and highlighted for context, the table of contents and incoming and outgoing references are located on the left top and bottom respectively, and context information is placed on the right. Many specific visual elements are annotated. } %
\end{figure}

\begin{figure}
	\includegraphics[width=.8\textwidth]{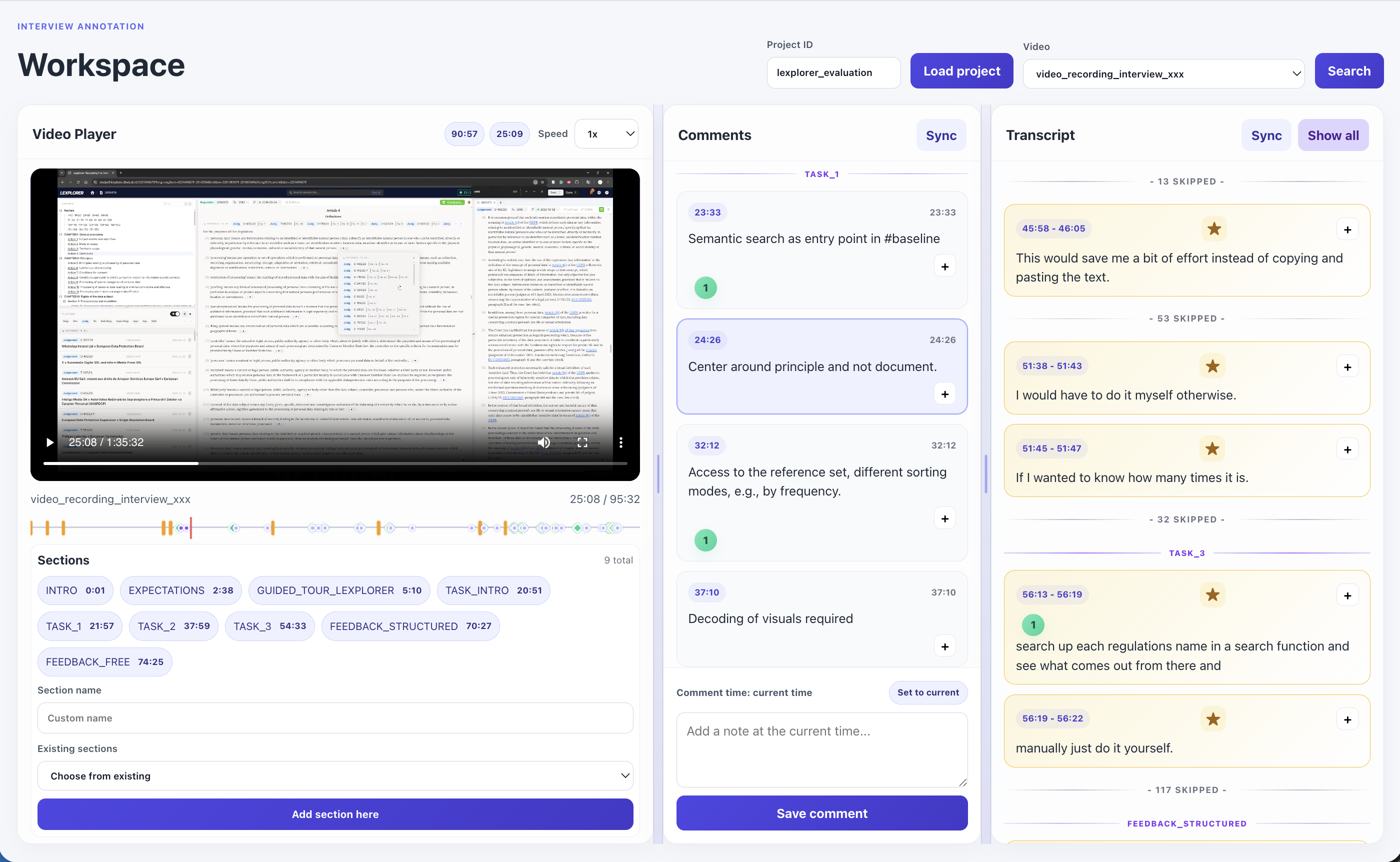}
	\caption{Screenshot of the custom annotation software we used for coding our interviews.}
	\label{fig:appendix-annotation-tool}
	\Description{A screenshot of the custom annotation software we used for coding our interviews.} %
\end{figure}

\clearpage

	\section{Interface Details}\label[appendix]{appendix:interface-details}

In the following, we describe the capabilities of \lexplorer in more detail, with a particular view to how they support meaning construction, 
elaborating on the summary provided in~\Cref{sec:interface-design}.

\subsection[Many-Documents View]{\lexManyDoc~~Many-Documents View}

The \emph{many-documents} view~(depicted in~\Cref{fig:many-documents-view}) stands at the beginning of legal inquiry.
By default, the view presents an overview of all the documents available in the corpus as a list of cards, allowing scholars to orient themselves and identify candidate sources for further inspection.
In the text mode~(\lexTextMode), each document is represented by its metadata, including short and long title, date, document type, unique identifier, authoring bodies, and keyword tags.
In the data mode (\lexDataMode), documents are represented as  glyphs~(see~\Cref{appendix:interface-design:one-document}). 
To narrow the scope of documents, users can combine metadata filters on document attributes with both instant meta-data-based search and full-text search, supporting both targeted retrieval and broader exploratory filtering.
Users can collect documents in dossiers to create sets of sources they would like to investigate together.
In text mode, the many-document view focuses on the selection and retrieval of documents (see~\Cref{fig:many-documents-view:text-mode}), whereas the data mode~(see~\Cref{fig:many-documents-view:data-mode}) exposes the relationships between the documents in a dossier via a force-directed layout.
Users can move to contextualized close-reading of individual documents or compare selections of documents from either~mode.

\subsection[One-Document View]{\lexOneDoc~~One-Document View}\label[appendix]{appendix:interface-design:one-document}

The \emph{one-document} view~(see~\Cref{fig:one-document-view}) supports contextualized close-reading of legal documents, while supporting lawyers in understanding \emph{incoming} and \emph{outgoing} influences after external changes.
By default, the view focuses on the text of the document considered with its structure as a table of contents~(ToC) on an upper-left pane.
The ToC allows readers to quickly situate themselves within the document to understand and navigate its formal legal structure.
Within the text of the document, interactive references enable navigation between structural entities within the document and navigation along relationships between documents.
This includes outgoing references to other documents and incoming ones, where other documents refer to the one under inspection.
The view embeds incoming references at a fine-grained level of resolution, e.g., at a letter of a paragraph in an article, allowing readers to see precisely which other documents target which parts of the current document.
This treatment of references ensures proper handling of normative intertextuality in law and draws on design from software engineering, where \emph{Go to definition} and \emph{Find usages} are two well-established, distinct operations with separate user interfaces.
Following outgoing relationships contextualizes the document within the landscape, while avoiding context switching by juxtaposing the relationships examined within the one-document view.
The tabbed arrangement of the side-by-side view enables legal scholars to follow legal traces such as chains of references without losing context.
In the text mode~(see~\Cref{fig:one-document-view:text-mode}), this view focuses on the close-reading of a document, while the data mode (\lexDataMode) emphasizes the relationships within the document's formal structure at a glance (see~\Cref{fig:one-document-view:data-mode}).

To achieve this, we use a glyph that builds on a modified sunburst visualization~\cite{staskoEvaluationSpacefillingInformation2000}, functioning as a visual ToC.
We choose this type of visualization since it has been previously used for document structure visualization~\cite{collinsDocuBurstVisualizingDocument2009}, participants in the requirement interviews have argued that their mental models rely on document hierarchies, and instances of the glyph can be connected to display legal relationships.
The glyph starts at 12 o'clock and progresses clock-wise, delimiting recitals from enacting terms if applicable.
The rings display the levels of the hierarchy with the upper-most level represented by the innermost ring.
The length of arcs is proportional to the text length of the corresponding structural entity.
We use the inner space of the glyph to give an overview of the internal references of a document, where orange ribbons aggregate these references between higher-order structural entities.
Similar to the aggregation of internal references, the data mode (\lexDataMode) also zooms out on relationships with other documents via aggregation to approximate the relevance of relations.

\subsection[Few-Documents View]{\lexFewDoc~~Few-Documents View}

The \emph{few-documents} view~(see~\Cref{fig:compare-view:text-mode,fig:compare-view:data-mode}) expands the capabilities of the one-document view, 
allowing legal scholars to interpret individual documents across languages and time~(\emph{comparison}), and relating different documents~(\emph{relation}).
For individual documents, the view juxtaposes different languages or versions next to each other, synchronizing their display to align on structural entities.
Alternatively, one can display changes in-line to directly see changed and unchanged parts, while the latter can also be collapsed.
Navigation within the text is still possible through scrolling as well as via the same ToC as presented in the one-document view.
This enables comparative legal analysis.
In particular, the comparison of different document versions in a specific language reveals differences with character-level granularity~(see~\Cref{fig:compare-view:text-mode:two-documents}), allowing in-depth analysis of legal changes.
Such analysis is usually only enabled by external tooling. 
\lexplorer allows the user to stay in context, avoiding mentally taxing context switching.
For different documents, their texts remain separate, allowing for individual inspection~(see~\Cref{fig:compare-view:text-mode:three-documents}).
Navigation of these documents is also still possible through separate ToCs.
The data mode (\lexDataMode) of the view emphasizes the relationships between different documents~(see~\Cref{fig:compare-view:data-mode:two-documents}).
Above the document's text, the data mode displays the same visualization as the one-document view but connects neighboring documents using directional edges that cross document panes.
The connections again follow the same fine-grained level of resolution as the one-document view and resolve chains of references in three-way comparisons~(see~\Cref{fig:compare-view:data-mode:three-documents}), 
enabling detailed analyses of the interplay between legal documents.
For ease of navigation, the visual display of relationships is linked with the documents' texts, enabling parallel navigation.

\clearpage

	\section{Prototype Evaluation Details}\label{appendix:prototype-evaluation}

In the following, we elaborate on our prototype evaluation discussed in \Cref{sec:prototype-evaluation},
providing the evaluation structure (\Cref{appendix:evaluation-structure})
as well as an in-depth overview of coded participant statements and feedback  (\Cref{appendix:prototype-evaluation-results}).
For the detailed participant information, see the combined \Cref{tab:participant-demographics} in \Cref{appendix:participant-demographics-table}, referring to the Prototype Evaluation (PE) column.

\subsection{Evaluation Structure}\label[appendix]{appendix:evaluation-structure}

Each evaluation consisted of four main parts.
In an introductory part,
we collected information about the participants
and
asked about expectations with regard to certain features.
After a quick introduction to the baseline (\emph{EurLex})
and
\lexplorer,
participants were prompted to complete three exploratory parts,
each relating to another perspective on the legal document landscape.
The evaluation was wrapped up by a quantitative feedback form
and a short semi-structured interview on usability and efficiency.

\subsubsection{Welcome, Demographics, Expectations \textit{(15 minutes)}}
\paragraph{1. Technical Setup}
\begin{itemize}
	\item Hardware-related:
	\begin{itemize}
		\item What screen size do you use?
		\item Do you use a mouse or a trackpad?
	\end{itemize}
	\item Software-related:
	\begin{itemize}
		\item What operating system do you use?
		\item What browser do you use?
	\end{itemize}
\end{itemize}
\paragraph{2. Demographics}
\begin{itemize}
	\item What is your participant ID? \textit{(assigned; name held separately on the consent form)}
	\item What is your age? \textit{(exact / 5-year-bands / prefer not to disclose)}
	\item Which gender do you identify with? \textit{(female / male / non-binary / prefer to self-describe / prefer not to disclose)}
	\item What group do you consider yourself to belong to? \textit{(scholarly / societal / institutional)}
	\item What are your previous and current roles? \textit{(free text)}
	\item How many years of experience do you have \dots
		\begin{itemize}
			\item \dots in general? \textit{(exact)}
			\item \dots with European legal documents? \textit{(exact)}
			\item \dots in using data visualizations or interactive visualizations? \textit{(exact)}
		\end{itemize}
	\item Please answer the following questions on a scale of 1 to 5 (\emph{not at all familiar}, \emph{slightly familiar}, \emph{somewhat familiar}, \emph{moderately familiar}, \emph{extremely familiar}): How familiar do you feel with the EU legal area of \dots
	\begin{itemize}
			\item \dots digital law.
			\item \dots sustainability law.
			\item \dots migration law.
		\end{itemize}
\end{itemize}

\paragraph{3. Expectations}

Think of a software prototype designed to help people who work with (European Union) law navigate and understand (EU) legal documents more quickly and comfortably.
In that context,
please answer the following questions on a scale from 1 to 5,
where the options are: 
\emph{not at all important},
\emph{slightly important},
\emph{somewhat important},
\emph{moderately important},
and \emph{extremely important}.
[\emph{Note: These statements were randomized in order per-group.}]

\begin{enumerate}[label=(\alph*)]
	\item Document Search \& Navigation
	\begin{itemize}
		\item How important is the ability to search for information within a legal document to you?
		\item How important is the ability to navigate between related legal documents for you?
	\end{itemize}
	\item Document Comparison \& Versions
	\begin{itemize}
		\item How important is the ability to compare different versions of the same legal document (e.g., amendments) for you?
		\item How important is the ability to understand how a legal document has changed over time for you?
	\end{itemize}
	\item Relationships \& Structure
	\begin{itemize}
		\item How important is the ability to identify how legal documents relate to one another (e.g., citations, amendments, dependencies) for you?
		\item How important is the ability to understand the structure of a single legal document to you?
	\end{itemize}
	\item Interface Features
	\begin{itemize}
		\item How important is a visual representation of document structure and relationships to you?
		\item How important are short content loading times when working with legal documents to you?
	\end{itemize}
\end{enumerate}

Are there any other expectations that you would like to share with us? (\textit{free text})

\subsubsection{Warm-Up \textit{(10 minutes)}}
\begin{itemize}
	\item Guided Tour of EUR-Lex
	\item Guided Tour of \lexplorer
\end{itemize}
\subsubsection{Quantitative Evaluation \textit{(45 minutes)}}

\paragraph{Introduction:}
\begin{quote}
	In the following 45 minutes, you will explore the Lexplorer prototype in the context of three different legal domains. 

Before starting the exploration in any specific context, we will ask you to comment on how you would approach it in the systems you would normally use (e.g., EUR-Lex or CURIA), and what might be challenging about doing so.

As part of the exploration using the prototype, we will guide you through the different features of our prototype and will ask you to comment on how you would imagine using these features in your own work. 

You do not need to know anything about the example domains--we are interested in learning about how you interact with our prototype, regardless of your level of expertise.
Remember that we are evaluating our prototype, not your performance.

\end{quote}

\paragraph{Task 1--Definitions and Principles in Digital Law \textit{(15 minutes)}}
\begin{quote}
The key legal concepts in any field of law are often grounded in legislative definitions and principles whose meanings can change over time, e.g., through judicial interpretation. 

To understand the current meaning of a legislative definition, one often needs to consider the case law engaging with that definition. For example, the highest court in the European Union (i.e., the Court of Justice) has commented extensively on the notion of personal data, a key concept in EU digital law defined in Article 4(1) of the General Data Protection Regulation.

In the following 15m, you will have a chance to explore how the notion of personal data has been shaped by the Court of Justice using the Lexplorer system. 

While you are exploring, please talk us through what questions you would like to ask and what you would like to achieve in the system. If needed, we will help you locate the relevant functionality, which may go beyond the functionality of the systems you are used to, in our prototype.
	
\end{quote}
\paragraph{Task 2--Goals and Actions in Sustainability Law \textit{(15 minutes)}}
\begin{quote}
While the meaning of legal provisions is organically shaped through judicial interpretation, legislative amendments can introduce abrupt changes. 

For example, the European Climate Law, a key instrument in EU sustainability law, originally contained many lofty aspirations, which were later turned into more tangible commitments. 

In the following 15m, you will use Lexplorer to explore how the EU's sustainability commitments, specifically those featured in the European Climate Law, have evolved over time, and how the European Climate Law interacts with some other instruments in the area of sustainability law. 

While you are exploring, please talk us through what questions you would like to ask and what you would like to achieve in the system. If needed, we will help you locate the relevant functionality, which may go beyond the functionality of the systems you are used to, in our prototype.

\end{quote}
\paragraph{Task 3--Order and Chaos in Migration Law \textit{(15 minutes)}}
\begin{quote}
While areas of law often take shape organically, some areas of law are (re)constituted by groups of legal instruments designed to form a coherent whole. 

A recent example can be found in the Common European Asylum System (CEAS), a milestone in EU migration law comprising, inter alia, Regulations 2024/1347, 2024/1348, 2024/1351, 2024/1358, and 2024/1359. 

In the following 15m, you will have a chance to explore the interplay between the different CEAS instruments mentioned above using the Lexplorer prototype. 

While you are exploring, please talk us through what questions you would like to ask and what you would like to achieve in the system. If needed, we will help you locate the relevant functionality, which may go beyond the functionality of the systems you are used to, in our prototype.
\end{quote}

\subsubsection{Qualitative Evaluation and Wrap-Up \textit{(20 minutes)}}

\paragraph{1. Efficiency \& Usability}
\begin{enumerate}[label=(\alph*)]
	\item For the prototype, please rate your agreement with the following statements based on a scale of 1 to 5, where the options are: \emph{strongly disagree}, \emph{disagree}, \emph{neutral}, \emph{agree}, and \emph{strongly agree}:
	\begin{itemize}
		\item I was able to quickly locate documents relevant to my legal tasks.
		\item I was able to explore the context and relationships of a given legal document.
		\item I was able to compare different languages and versions of the same legal document.
		\item I was able to easily keep track of the documents relevant to my legal tasks.
	\end{itemize}
	\item For the prototype, please rate your agreement with the following statements based on a scale of 1 to 5, where the options are: \emph{strongly disagree}, \emph{disagree}, \emph{neutral}, \emph{agree}, and \emph{strongly agree}:
	\begin{itemize}
		\item Accomplishing my legal tasks required little mental effort. (\textit{reverse})
		\item I am satisfied with how effectively I completed my legal tasks.
		\item I had to rely on trial and error to figure out how to use the interface. (\textit{reverse})
		\item The interface felt intuitive and responsive.
	\end{itemize}
\end{enumerate}

\paragraph{2. Inquiry on Efficiency \& Usability}

The following questions depended on the answers of the participant to the previous questions and were only posed if applicable:
\begin{enumerate}[label=(\alph*)]
	\item What made it easy/hard for you to find the document relevant to your legal tasks, and why?
	\item What made it easy/hard for you to explore the context of a given legal document, and why?
	\item What made it easy/hard for you to compare different languages and versions of the same legal document, and why?
	\item What made it easy/hard for you to keep track of the documents relevant to your legal tasks, and why?
	\item Why did accomplishing your legal tasks require little/lots of mental effort?
	\item Why were you satisfied/dissatisfied with your performance?
	\item Why did you have to/not have to rely on trial-and-error to figure out how to use the interface?
	\item What made the interface feel intuitive or responsive/unintuitive or unresponsive?
\end{enumerate}

\paragraph{3. Open-Ended Questions}
\begin{enumerate}[label=(\alph*)]
	\item If you had to perform a complex legal research task for your actual work or studies, and assuming that all relevant legal data would be available in Lexplorer, which system would you choose, and why?
	\item Did you find added value in using the prototype compared to the baseline?
	\begin{itemize}
		\item \textit{Probe}: Did you make any new discoveries about legal documents that you had already worked with before the evaluation?
	\end{itemize}
	\item Were there any questions that came up during the legal tasks that you could answer in one system but not in the other system, or questions you could not answer using either system?
	\item What would you change about the prototype, and why?
	\begin{itemize}
		\item \textit{Probe}: If you had a magic wand, what features would you want to add to the prototype and why?
	\end{itemize}
\end{enumerate}

\paragraph{4. Conclusion}

\begin{itemize}
	\item Is there anything else related to this interview that you would like to share with us?
	\item Is there someone else you would recommend who might be interested in evaluating the prototype with us?
\end{itemize}

\subsection{Detailed Evaluation Results}\label[appendix]{appendix:prototype-evaluation-results}

In \Cref{tab:evaluation-prototype-results}, we present our findings
from the qualitative parts of the prototype evaluation in greater detail,
providing an index of coded statements,
corresponding participants,
and code frequencies.
\clearpage

\begin{longtblr}[
  caption = {Overview of the 62 codes extracted from the qualitative parts of the Prototype Evaluation,
  organized across 8 themes.},
  label = {tab:evaluation-prototype-results},
]%
{
	colspec={
		Q[l,wd=1.5cm]
		Q[l,wd=1.5cm]
		Q[l,wd=1.4cm]
		Q[l,wd=6cm]
		Q[l,wd=3.75cm]
		Q[r,wd=0.25cm]
	},
	column{1} = {leftsep=0pt, rightsep=0.25em},
	column{2} = {leftsep=0pt, rightsep=0.25em},
	column{3} = {leftsep=0pt, rightsep=0.25em},
	column{4} = {leftsep=0pt, rightsep=0.75em},
	column{5} = {leftsep=0pt, rightsep=0.25em},
	column{6} = {leftsep=0pt, rightsep=0pt}
}
\toprule
\textbf{Theme} & \textbf{Category} & \textbf{ID} & \textbf{Statement} & \textbf{Participants} & \textbf{\#} \\
\midrule
\textbf{Baseline} & Tools & B-BO & \dots uses \emph{Beck-Online} to access legal documents & $P_{22}$, $P_{29}$, $P_{25}$, $P_{5}$ & 4 \\
&  & B-BT & \dots opens documents in multiple tabs & $P_{21}$, $P_{11}$, $P_{13}$, $P_{19}$, $P_{20}$, $P_{22}$, $P_{17}$ & 7 \\
&  & B-C & \dots uses \emph{Curia} to access European jurisprudence & $P_{19}$, $P_{20}$, $P_{27}$, $P_{18}$, $P_{22}$, $P_{21}$, $P_{30}$ & 7 \\
&  & B-MSWC & \dots uses text-editing software like \emph{Microsoft Word} to compare versions & $P_{28}$, $P_{24}$ & 2 \\
& Browser & B-BCMDF & \dots uses in-browser functionality \emph{CMD+F} & $P_{17}$, $P_{23}$, $P_{29}$, $P_{20}$, $P_{9}$ & 5 \\
&  & B-G & \dots uses a general-purpose search engine like \emph{Google} to search for documents & $P_{25}$, $P_{22}$, $P_{18}$, $P_{19}$ & 4 \\
& Other & B-AC & \dots starts legal research with scholarly material & $P_{12}$, $P_{22}$, $P_{21}$, $P_{29}$, $P_{28}$, $P_{9}$, $P_{25}$, $P_{19}$ & 8 \\
\midrule
\textbf{Cognitive Load} &  & C-AC & \dots appreciated high number of functionalities & $P_{23}$, $P_{16}$, $P_{13}$, $P_{18}$, $P_{22}$, $P_{11}$ & 6 \\
 &  & C-LIC & \dots mentioned that law is inherently complex & $P_{13}$, $P_{5}$, $P_{9}$, $P_{29}$, $P_{25}$, $P_{22}$ & 6 \\
 &  & C-MOT & \dots showed confidence that familiarity with the prototype was only a matter of time & $P_{13}$, $P_{18}$, $P_{17}$, $P_{27}$, $P_{19}$, $P_{12}$, $P_{26}$, $P_{25}$, $P_{30}$ & 9 \\
 &  & C-O & \dots felt overwhelmed & $P_{16}$, $P_{12}$, $P_{9}$, $P_{23}$, $P_{17}$, $P_{18}$, $P_{27}$, $P_{20}$, $P_{25}$ & 9 \\
 &  & C-TAEN & \dots perceived trial-and-error as a `normal process' & $P_{17}$, $P_{19}$, $P_{29}$, $P_{5}$ & 4 \\
\midrule
\textbf{Data} & Coverage & D-DA & \dots asked to include administrative documents & $P_{5}$, $P_{24}$, $P_{22}$, $P_{21}$ & 4 \\
 &  & D-DC & \dots asked to include national documents & $P_{5}$, $P_{29}$, $P_{24}$, $P_{9}$, $P_{20}$ & 5 \\
 &  & D-DP & \dots asked to include procedural data & $P_{5}$, $P_{18}$, $P_{26}$, $P_{24}$ & 4 \\
 &  & D-IL & \dots asked to include international law & $P_{13}$, $P_{9}$, $P_{18}$, $P_{29}$ & 4 \\
 & Trust & D-AC & \dots asked for data completeness or accuracy & $P_{12}$, $P_{22}$, $P_{5}$, $P_{30}$, $P_{29}$, $P_{21}$, $P_{24}$ & 7 \\
\midrule
\textbf{Multi-Doc-Views} & Use Case & MD-AC & \dots described presentation of change as core of doctrinal legal research & $P_{29}$, $P_{12}$, $P_{22}$ & 3 \\
 &  & MD-UCLANG & \dots mentioned use case of comparing languages & $P_{29}$, $P_{28}$, $P_{19}$, $P_{18}$, $P_{30}$ & 5 \\
 &  & MD-UCRC & \dots mentioned use case of detecting recent changes to a document & $P_{23}$, $P_{11}$, $P_{9}$, $P_{28}$, $P_{20}$, $P_{18}$, $P_{29}$, $P_{26}$, $P_{25}$, $P_{22}$ & 10 \\
 &  & MD-UCT & \dots mentioned use case of tracking a legal concept over time & $P_{12}$, $P_{18}$, $P_{24}$, $P_{13}$, $P_{17}$, $P_{26}$, $P_{23}$, $P_{30}$ & 8 \\
 & Feedback & MD-ILC & \dots felt that language comparison was intuitive & $P_{18}$ & 1 \\
 &  & MD-ITC & \dots felt that temporal comparison was intuitive & $P_{24}$, $P_{17}$, $P_{29}$, $P_{26}$, $P_{25}$, $P_{22}$, $P_{11}$, $P_{30}$ & 8 \\
 &  & MD-MS & \dots uses comparison functionality of \emph{Microsoft Word} as baseline & $P_{16}$, $P_{13}$ & 2 \\
 &  & MD-SBS & \dots generally appreciated side-by-side layout & $P_{16}$, $P_{17}$, $P_{18}$, $P_{29}$, $P_{24}$, $P_{13}$, $P_{9}$, $P_{12}$, $P_{26}$, $P_{25}$, $P_{23}$, $P_{22}$, $P_{11}$, $P_{30}$ & 14 \\
 & Suggestions & MD-TF & \dots suggested to track changes over a longer period of time & $P_{13}$, $P_{19}$, $P_{24}$, $P_{17}$, $P_{26}$, $P_{23}$, $P_{18}$ & 7 \\
 &  & MD-TL & \dots suggested a timeline as visual aid & $P_{13}$, $P_{18}$ & 2 \\
\midrule
\textbf{References} & Use Case & R-CL & \dots mentioned case-law-analysis as a use case & $P_{12}$, $P_{22}$, $P_{11}$, $P_{29}$, $P_{20}$, $P_{9}$, $P_{26}$, $P_{25}$, $P_{21}$, $P_{19}$, $P_{24}$ & 11 \\
& Analysis & R-ADT & \dots analyzed references based on document type & $P_{12}$, $P_{13}$, $P_{5}$, $P_{25}$, $P_{22}$, $P_{29}$, $P_{20}$, $P_{9}$, $P_{26}$, $P_{21}$, $P_{19}$, $P_{24}$ & 12 \\
 &  & R-ALR & \dots analyzed references based on location of references & $P_{21}$, $P_{9}$, $P_{22}$, $P_{19}$ & 4 \\
 &  & R-ANR & \dots analyzed references based on number of references & $P_{25}$, $P_{21}$, $P_{19}$ & 3 \\
 &  & R-AT & \dots analyzed references based on date of the document & $P_{25}$, $P_{9}$ & 2 \\
 & Baseline & R-MP & \dots would normally parse references manually & $P_{16}$, $P_{5}$, $P_{20}$, $P_{26}$, $P_{25}$, $P_{21}$ & 6 \\
 &  & R-OT & \dots looks up reference targets in other tab(s) & $P_{20}$, $P_{22}$, $P_{11}$ & 3 \\
 &  & R-TS & \dots uses text-search to find relations and references & $P_{20}$, $P_{26}$ & 2 \\
 & Feedback & R-AR & \dots generally appreciated reference functionalities & $P_{16}$, $P_{5}$, $P_{28}$, $P_{20}$, $P_{9}$, $P_{18}$, $P_{22}$, $P_{21}$, $P_{24}$, $P_{11}$ & 10 \\
 &  & R-FG & \dots appreciated fine-grained quality of references & $P_{9}$, $P_{20}$, $P_{29}$, $P_{25}$ & 4 \\
 &  & R-IP & \dots felt that reference functionality increases productivity & $P_{13}$, $P_{29}$, $P_{22}$, $P_{17}$ & 4 \\
 &  & R-IT & \dots showed intuitiveness for in-text references & $P_{18}$, $P_{22}$, $P_{24}$, $P_{9}$, $P_{19}$ & 5 \\
 &  & R-R & \dots showed intuitiveness for references ribbons & $P_{19}$, $P_{18}$, $P_{29}$, $P_{24}$, $P_{20}$, $P_{25}$, $P_{22}$ & 7 \\
 &  & R-U & \dots felt that reference functionality was useful & $P_{12}$, $P_{18}$, $P_{24}$, $P_{13}$, $P_{9}$, $P_{26}$, $P_{25}$, $P_{22}$, $P_{19}$ & 9 \\
 & Suggestions & R-WOR & \dots wished for custom re-ordering of reference sets & $P_{16}$, $P_{13}$, $P_{26}$ & 3 \\
 &  & R-WRC & \dots wished for reference chains & $P_{18}$, $P_{17}$, $P_{25}$ & 3 \\
 &  & R-WTS & \dots wished to query the reference set via text & $P_{18}$, $P_{17}$, $P_{26}$ & 3 \\
\midrule
\textbf{Semantical Thinking} & Text & ST-IDTS & \dots searched within a document via text query & $P_{19}$, $P_{23}$, $P_{17}$, $P_{27}$, $P_{9}$ & 5 \\
 &  & ST-T & \dots relied on short titles or abbreviations & $P_{13}$, $P_{18}$, $P_{21}$, $P_{11}$, $P_{24}$, $P_{27}$, $P_{20}$, $P_{9}$, $P_{25}$ & 9 \\
 &  & ST-TC & \dots pointed to terminological distinction between change within a document and amendments made to other documents & $P_{29}$, $P_{19}$ & 2 \\
 &  & ST-TS & \dots approached a task trough text search & $P_{16}$, $P_{27}$, $P_{18}$, $P_{25}$, $P_{21}$ & 5 \\
 & Feedback & O-A & \dots appreciated the interactive table of content & $P_{22}$, $P_{16}$, $P_{9}$, $P_{18}$, $P_{21}$ & 5 \\
 &  & ST-CC & \dots appreciated the color coding & $P_{29}$, $P_{17}$, $P_{5}$ & 3 \\
 &  & ST-DM & \dots asked to include document-type-specific metadata & $P_{21}$, $P_{26}$, $P_{19}$, $P_{24}$ & 4 \\
 &  & ST-DT & \dots showed active awareness for different document types & $P_{12}$, $P_{13}$, $P_{27}$, $P_{29}$, $P_{26}$, $P_{25}$, $P_{21}$ & 7 \\
 &  & ST-ST & \dots felt supported in systematic legal thinking & $P_{29}$ & 1 \\
 \midrule
\textbf{Tasks} & Feedback & T-G & \dots felt that the tasks were too general & $P_{12}$, $P_{19}$ & 2 \\
 \midrule
\textbf{Visual Components} & Experience & V-LI & \dots reported little imagination for visualizations & $P_{16}$, $P_{17}$, $P_{19}$, $P_{25}$, $P_{28}$ & 5 \\
 &  & V-NE & \dots reported no/little experience with visualizations & $P_{16}$, $P_{21}$, $P_{23}$ & 3 \\
 & Feedback & V-NSV & \dots did not struggle to navigate via the glyph visualization & $P_{18}$, $P_{28}$, $P_{29}$, $P_{22}$, $P_{17}$, $P_{11}$ & 6 \\
 &  & V-SV & \dots struggled to navigate via the glyph visualization & $P_{27}$, $P_{19}$, $P_{21}$, $P_{13}$, $P_{30}$ & 5 \\
 &  & V-UI & \dots found useful information in the glyph visualization & $P_{29}$, $P_{9}$, $P_{26}$, $P_{17}$, $P_{24}$ & 5 \\
 &  & V-NUI & \dots did not find useful information in the glyph visualization & $P_{27}$, $P_{20}$, $P_{25}$, $P_{23}$, $P_{22}$ & 5 \\
 &  & V-VA & \dots generally appreciated the visual components & $P_{16}$, $P_{29}$, $P_{26}$, $P_{17}$, $P_{24}$, $P_{11}$ & 6 \\
 & Suggestions & V-WTL & \dots wished for a list of references in combination to the glyph & $P_{21}$, $P_{13}$, $P_{19}$, $P_{30}$ & 4 \\
\bottomrule
\end{longtblr}

\clearpage

\end{document}